\documentstyle[prd,aps,floats]{revtex}
\input{psfig.tex}

\tighten
\begin{document} 
\draft
\title{Planck-scale quintessence and the physics of structure formation}
\author{Constantinos Skordis and Andreas Albrecht}
\address{Department of Physics, UC Davis, One Shields Ave, Davis CA 95616}
\maketitle
\begin{abstract}

In a recent paper we considered the possibility of a scalar field
providing an explanation for the cosmic acceleration. Our model had the
interesting properties of attractor-like behavior and having its 
parameters of $O(1)$ in Planck units. Here we discuss the effect of
the field on large scale structure and CMB anisotropies.  We show how
some versions of our model inspired by ``brane'' physics have 
novel features due to the fact that the scalar field has a 
significant role over a wider range of redshifts than for typical
``dark energy'' models.  One of these features is the additional
suppression of the formation of large scale structure, as compared with 
cosmological constant models.  In light of the new pressures being
placed on cosmological parameters (in particular $H_0$) by CMB
data, this added suppression allows our ``brane'' models to give excellent fits to
both CMB and large scale structure data.
\end{abstract}

\date{\today}


\renewcommand{\thefootnote}{\arabic{footnote}}
\setcounter{footnote}{0}

\section{Introduction}
Current evidence that the expansion of the Universe is accelerating\cite{BOPS},
if confirmed, requires dramatic changes in the field of theoretical
cosmology.   Until recently, there was strong prejudice against the
idea that the Universe could be accelerating.  There simply is no
compelling 
theoretical framework that could accommodate an accelerating universe.
Since the case for an accelerating universe continues to build (see
for example \cite{JaffeEtal:00}), 
attempts have been made to improve the theoretical situation, with
some modest success.  Still, major ``fine tuning'' problems remain.

All attempts to account for acceleration [3-31]
introduce a new type of matter (the ``dark energy'' or ``quintessence'')
with an equation of state $p_\phi = w_\phi \rho_\phi$ relating pressure and
energy density.  Values of $w_\phi \leq -0.6$ today are
preferred by the data\cite{PTW} and in many models $w_\phi$ can vary over time.  (In this framework, a
plain cosmological constant is the case where $w_\phi= -1$ independent of
time.) Most models with a varying $w_\phi$ are based on a simple scalar field with 
a specific potential in the Lagrangian. In addition to the standard ingredients fed into 
the  Friedman-Robertson-Walker (FRW) cosmology, one introduces a scalar field which in
general has a varying $w_\phi$. If the value of $1 + 3W_\phi\Omega_\phi$ 
becomes negative (with the assumption that
the photon and neutrino contributions to the total density of the universe is small compared to
the rest of the mass-energy) the universe enters an era of accelerated expansion.
 
The evolution of spatially homogeneous scalar fields in an FRW
cosmology has a long history in the context of cosmic inflation, but
the inflaton fields do not play a significant role today. In the last two years 
scalar fields have been considered which can produce an accelerated
expansion in the present epoch
This area has been stimulated by the growing evidence 
for cosmic acceleration today.  The additional scalar field matter is
known as ``quintessence'' or ``dark energy''. In the early models one had to fine-tune 
the initial conditions of the field in order to get an accelerated expanding universe 
today, a feature which is very undesirable. Later on the very interesting category of 
``tracker'' quintessence models were created, in which the field gives the desired 
behavior independently of initial conditions. Still one had to introduce a small scale 
into the Lagrangian in order to achieve this.  One way forward with
these models is to try and construct a specific explanation for these small
parameters~\cite{PBBM,BM}.  
  
  In a recent paper\cite{AS} we discussed a class of quintessence models which like 
the ``tracker'' models, would produce the desired effect of an accelerated universe, 
independently of the initial conditions. The model was based on the pure exponential 
potential which was known to possess attractor
solutions\cite{H,B,CW,RP,CLW,BCC,FJ}. However, the pure exponential 
however, in the attractor regime, cannot produce a realistic
accelerated expanding universe. In our previous  paper we showed
 that when the exponential potential is modified by a polynomial prefactor, we can keep
 the attractor solutions during most of the history of the universe (and therefore the 
independence from the initial conditions) while at the same time producing an accelerating
 scale factor today. The new nice feature of the model, not found previously in other 
models, was that all the parameters involved were $O(1)$ in Planck
units.  Since then, a number of authors taken this idea in interesting
directions \cite{ArkaniHamedEtal:00,BeanMagueijo:00}, which adds to
the case for pursuing the observable effects of our model on cosmic structure.
(We also note that yet another class models using simple Planck-scale
physics have been proposed \cite{Wetterich:00}).

In this paper
we consider the evolution of small perturbations for the former potential 
and discuss the implications
on the Cosmic Microwave Background (CMB) and  structure formation. 
In particular the recent Boomerang and Maxima
data suggest that the Hubble constant might be larger than the so far 
accepted value of $65kms^{-1}Mpc^{-1}$.
In this case if one normalizes the data to COBE, 
the standard $\Lambda$-CDM model doesn't do that well for the observed matter power spectrum. 
 The former potential seems to do slightly 
better. One of our important results comes from a modification to the
potential reported in \cite{AS}, inspired by ``brane'' physics which
results in a number of novel features.  One of these features is a
much better joint fit to the matter power spectrum and CMB anisotropy power. 
 
The structure of this paper is as follows. First we review the evolution of scalar fields  in an
FRW cosmology. We discuss in particular the two mentioned potentials
(the one from \cite{AS} and the ``brane'' variation) emphasizing the relevant physics and
noting the differences between the two of them and between other models. Then we present the evolution 
of small perturbations for the two fields at hand and discuss the various effects. The evolution of
perturbations is different in the two models which leads to considerable differences in the power spectra. 
The effect of the additional dark energy in the universe on the matter power spectrum is discussed next. We
consider parameters of the models that give good $\sigma_8$'s and fit
the newly published decorrelated 
data of Hamilton et.al.~\cite{HT,HTP}. Following that we focus on the related physics
that affect the CMB and also discuss
the difference between the two models and $\Lambda$ models. A final
section gives our conclusions.

 \section{Evolution of the background}
Let us review the background evolution for a scalar field cosmology. This section does 
not introduce anything new (except the final part which introduces the new potential) and 
is merely intended for the new readers not familiar with the subject. We use a $(+---)$ 
metric signature, with Greek indices running from $0$ to $3$ and Latin indices (spatial) 
from $1$ to $3$. Index free $3$-vectors are denoted as $\vec{x}$ and $4$-vectors as
$\underline{x}$. Unless otherwise told, all units will be Planckian
with $M_p= \frac{1}{\sqrt{k_e}}\equiv (8\pi
G)^{(-1/2)}=1$ , $k_e$ being Einstein' constant and $M_{p}$ being the
Planck mass. Moreover we assume that the  
zeroth order cosmology is flat FRW in the conformal synchronous gauge.

\subsection{Cosmological scalar field evolution}
The point of departure is the action functional for the scalar field and gravity,
\begin{equation} 
	S[g_{\alpha\beta},\phi] = \int d^4x \sqrt{-g}\{\frac{R}{2} - \frac{1}{2}\phi_{,\mu}\phi^{,\mu} + V(\phi)\}
	\label{eq:genaction}
\end{equation}
Since the background metric has been fixed by our assumptions we can use 
$\sqrt{-g} = a^4$ and $R = -6\frac{\ddot{a}}{a^3}$.
Moreover the assumption of homogeneity and anisotropy forces the spatial derivatives of the field to zero,
$\phi_{,i}=0$.
The action simplifies, and takes the form of one particle Lagrangian mechanics in two dimensions with
coordinates $a(\tau)$ and $\phi(\tau)$.
\begin{equation}
   S[a(\tau),\phi(\tau)] = \int d\tau \{ 3\dot{a}^2 - \frac{1}{2}a^2\dot{\phi}^2 + a^4V(\phi)\} 
	\label{eq:action}
\end{equation}
The canonical momenta are $p_a = 6\dot{a}$ and $p_\phi = a^2\dot{\phi}$. The constraint $H=0$  gives
the first Friedman equation
\begin{equation}
	3(\frac{\dot{a}}{a^2})^2 = \frac{1}{2a^2}\dot{\phi}^2 + V(\phi)
	\label{eq:Friedman1}
\end{equation}
and therefore the field density is
\begin{equation}
	\rho = \frac{1}{2a^2}\dot{\phi}^2 + V(\phi)
	\label{eq:rhophi}
\end{equation}
Variation w.r.t. the field $\phi(\tau)$ gives the field equation of motion
\begin{equation}
	\ddot{\phi} + 2\frac{\dot{a}}{a}\dot{\phi} + a^2V_{,\phi} = 0
	\label{eq:backtemp}
\end{equation}
where $()_{,\phi}$ denotes derivative with respect to $\phi$.

Finally variation w.r.t. the scale factor $a(\tau)$ gives the second Friedman equation
\begin{equation}
 	6\frac{\ddot{a}}{a^3} = -\frac{1}{a^2}\dot{\phi}^2 +4V =  \rho - 3P
	\label{eq:Friedman2}
\end{equation}

from which we get  the field pressure as
\begin{equation}
	P = \frac{1}{2a^2}\dot{\phi}^2 - V(\phi)
	\label{eq:Pphi}
\end{equation}

If we want our scalar field to accelerate the universe we need the field to provide the dominant 
form of energy density and at the same time have negative pressure. In order to accelerate 
the universe a necessary and sufficient condition is that the 
deceleration parameter $q$ given by
\begin{equation}
	q  = - \frac{a \ddot{a}}{\dot{a}^2}= \frac{1}{2}(1 + 3w_\phi\Omega_\phi + \Omega_r )
\end{equation}
be negative or equivalently $P_\phi  < -H^2$ (In the last equation the dot denotes a derivative 
with respect to real time).

\subsection{The pure exponential potential}
One very interesting potential is the exponential $V=V_0 e^{-\lambda\phi}$. This potential 
has been shown to have attractor solutions~\cite{H,B}, that is regardless of initial conditions, the field
eventually scales like the dominant matter component. For a detailed discussion see~\cite{FJ}.

 If the dominant component scales as $\rho_n = \rho_0 (\frac{a_0}{a})^n$ then
 the scalar field approaches an attractor solution and its
fractional density is given by $\Omega_\phi = \frac{n}{\lambda^2}$. This is a special case of scaling
a field - a classification of such fields can be found in~\cite{LidSch}.

\begin{figure}[h]
\centerline{\hbox{\psfig{file=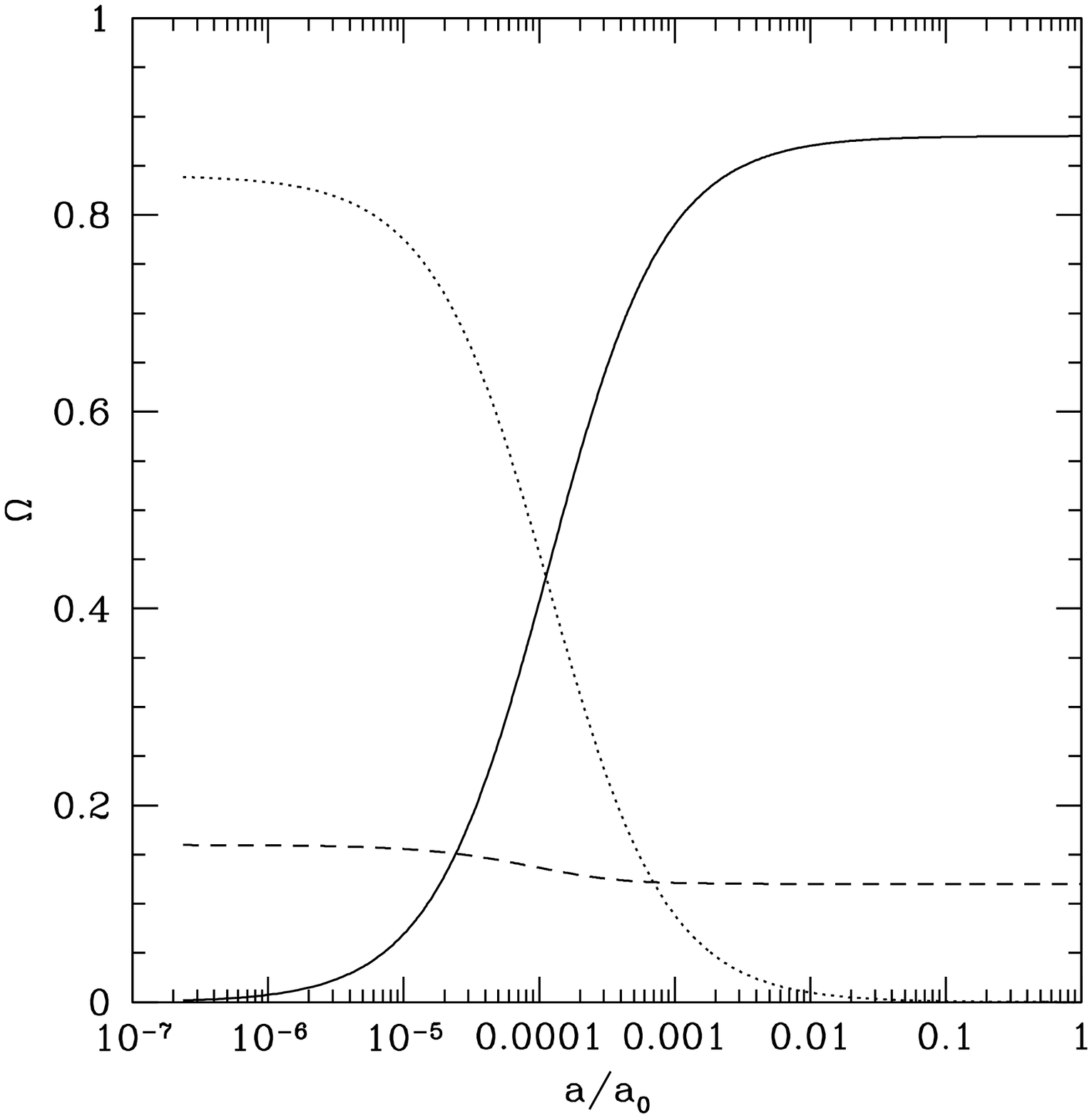,width=3.5in}\hspace{1cm}
\psfig{file=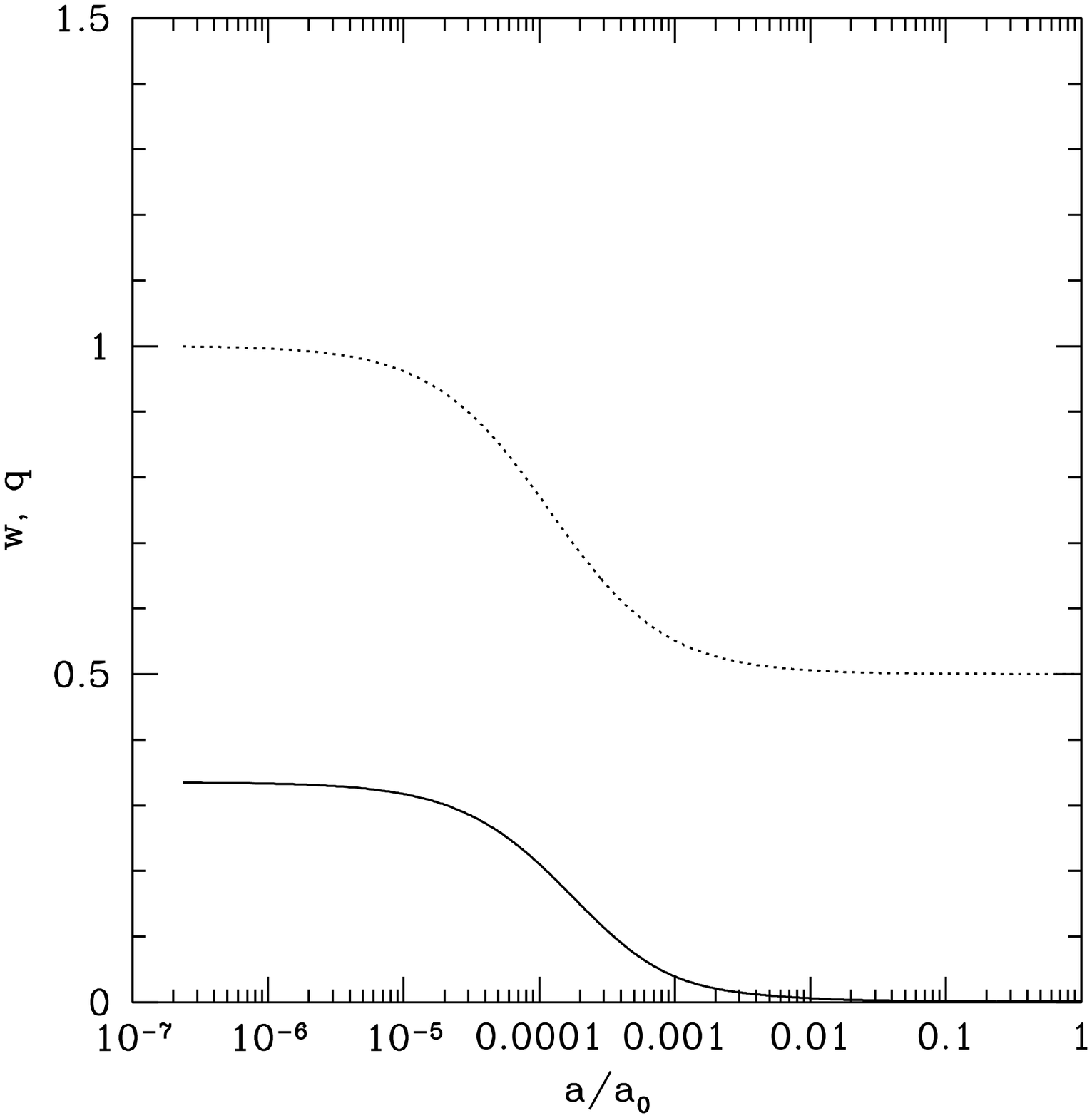,width=3.5in}}}
\caption{ 
{\bf LEFT:} Fractional densities $\Omega$ as a function $a$ for the
exponential potential. 
   The solid curve corresponds to matter with
  $\Omega_m\approx 0.88$ today, the dotted curve corresponds to radiation 
  $T=2.726K$ today and the dashed curve
  corresponds to the scalar field with $\Omega_\phi \approx 0.12$($\lambda=5$) 
  today. Radiation-to-matter transition is
  around $a=10^{-4}$ and $a=1$ today. \protect\newline
{\bf RIGHT:} Evolution of the equation of state parameter $w_{\phi}$ 
  and the deceleration parameter $q_\phi$
  as a function of a for the same model.
 The solid curve corresponds to $w$ and the dotted one to $q$. The
attractor causes $w_{\phi}$ to simply mimics the $w$ of the dominant
form of matter so $q_\phi$ can never be negative 
  in the attractor regime of a pure exponential potential.}
\label{expo}
\end{figure}

During the attractor era the relevant physical quantities are
\begin{eqnarray}	
	\rho_\phi = \frac{6}{6 - n}V = \frac{3}{na^2}\dot{\phi}^2 \\
	\frac{\dot{a}}{a} = (\frac{2}{n-2})\frac{1}{\tau}
	\label{eq:att_quant}
\end{eqnarray}
where $n = 3(1+w)$.

Figure \ref{expo} shows the evolution of
such a field with the scale factor. Notice the change of the fractional density
in the field from the radiation to the matter era. Since $w$ simply
mimics the $w$ of the dominant matter, it is 
impossible to get an accelerating universe in the attractor era. It is
possible to avoid the attractor if one chooses $\lambda < \sqrt{n}$. In that 
case the attractor is not there but the field will have its own scaling 
behavior with $w_\phi = \frac{\lambda^2}{3} -1$.   It is also
possible to choose initial conditions sufficiently far from the
attractor solution that the attractor has not been reached even today\cite{Huterer:99}.

\subsection{Modifying the exponential potential} 
The attractor behavior of the exponential is appealing, and it would
be nice if we could modify it in such a way as to keep this behavior for 
most of the history of the universe and at the same time get an accelerated 
universe today. Indeed we showed in a previous paper~\cite{AS} that this is 
possible by including a prefactor in front of the exponential $V=V_p(\phi)e^{-\lambda\phi}$.
 The effect of the prefactor is to introduce a local minimum in the
exponential such that the field gets trapped into it. After the field 
gets trapped it starts behaving like a cosmological constant and the 
universe eventually enters an era of accelerated expansion. In our
previous paper we used a polynomial 
prefactor. The full potential then takes the form 
\begin{equation}
	V = V_0[(\phi - B)^2 + A]e^{-\lambda \phi}.
	\label{eq:ASpotential}
\end{equation}
Through this paper we refer to this potential as the ``AS'' model. 

\begin{figure}[h]
\centerline{\hbox{\psfig{file=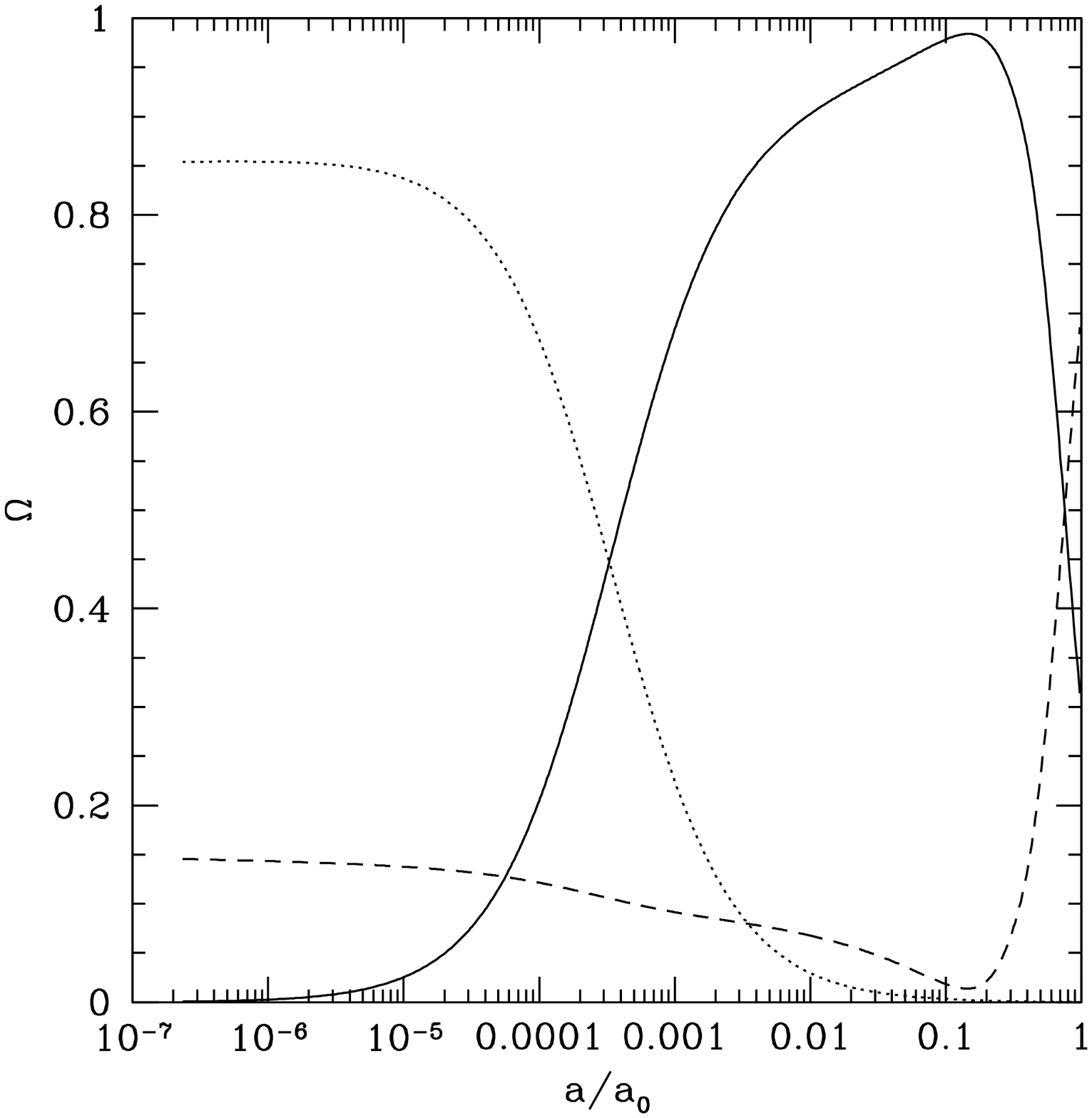,width=3.5in}\hspace{1cm}
\psfig{file=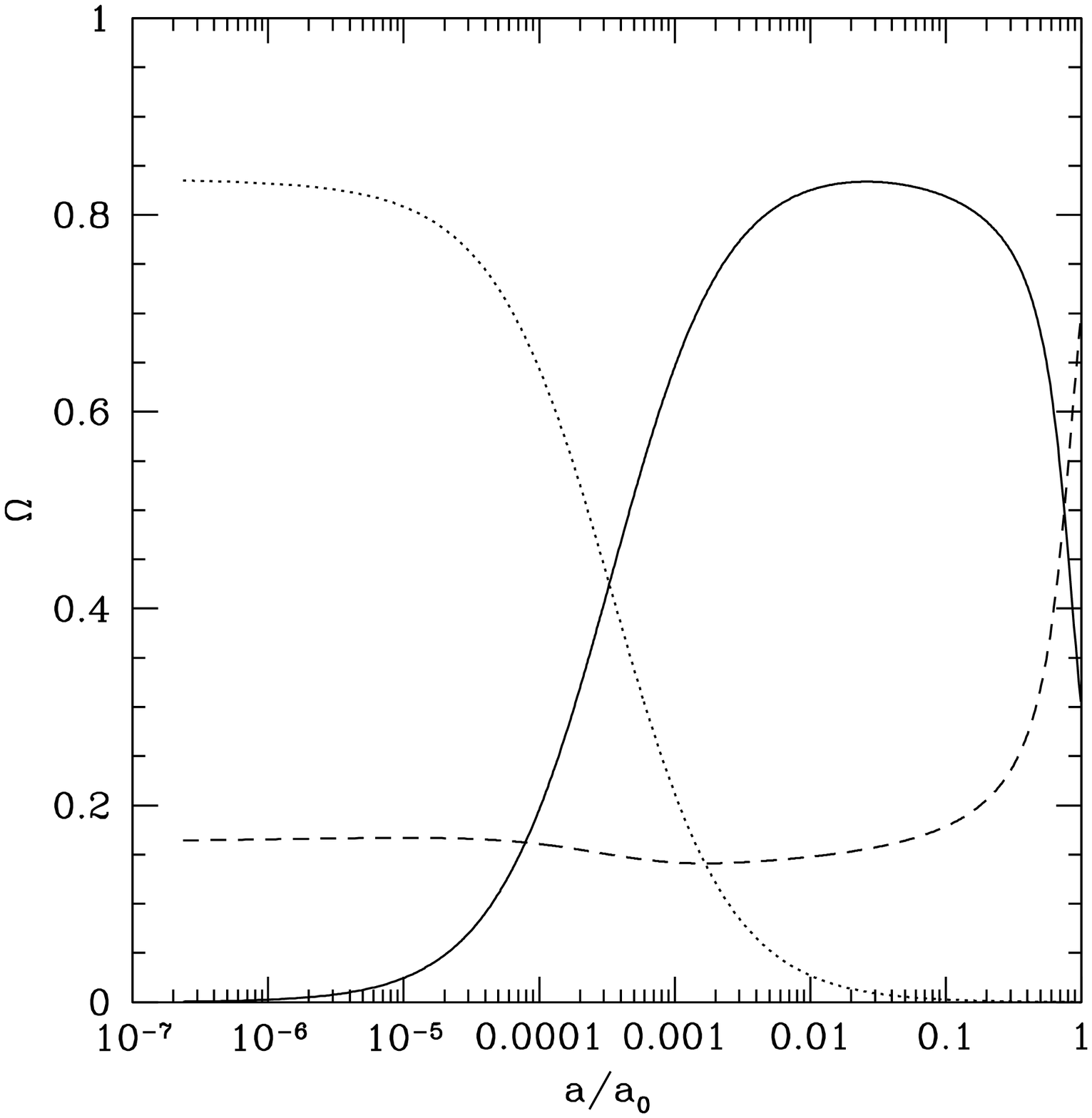,width=3.5in}}}
\caption{Solutions obtained by including a $V_p$ factor in
the potential. We show $\Omega_\phi(a)$ (dashed), $\Omega_{\rm \it matter}(a)$
(solid) and $\Omega_{\rm \it radiation}(a)$ (dotted). Radiation-to-matter transition is
  around $a=10^{-4}$ and $a=1$ today. Both models have $h = 0.65$, $\Omega_\phi =0.7$ and 
$\Omega_m = 0.247$ today.\protect\newline
{\bf LEFT:} The AS model  
with $\lambda=5$, $B=54.4057$, $A = .01$. Notice the behavior of the field at late times where
$\Omega_\phi$ goes very near to zero. That is the period of structure formation so the field
doesn't affect structure formation much.\protect\newline
 {\bf RIGHT:} The Brane model with $\lambda=5$, $B=56.10425$, $A = .01$, $C=1$ and $D=0.1$.
 Notice that the behavior of the field at late times is different from the AS model in the sense
that the field retains a significant amount of energy density. This
behavior affects structure formation as well as the CMB.}
\label{twopotentials}
\end{figure}

The left panel of Fig. \ref{twopotentials} shows the cosmological evolution of
densities in an AS model. One noticeable feature of this
model is the breakaway from the attractor behavior at late times. This will be important
for structure formation as the field density becomes very small when structure begins to form and
so it doesn't affect structure formation much. Lastly we should note
that trapping takes place independently of initial conditions of the
quintessence field in the very early Universe. This is
because the Universe is drawn into the attractor at early times.
Since the field is in the attractor solution by the time it approaches
the local minimum there is no memory of the initial 
conditions.  There is a variety of possible behaviors at the local
minimum. If the minimum is very shallow however the field can roll
through rather than be trapped. It is also possible 
to choose parameters which remove the local minimum, but the field
still lingers long enough to cause an era
of accelerated expansion.  For trapping a necessary condition
is $A\lambda^2<1$. The various behaviors and their corresponding
parameter ranges have
been investigated in ~\cite{BBM}. 

Instead of choosing a polynomial prefactor it is possible to have the
desired accelerated  expansion by using a different function
$V(\phi)$. An oscillating term for example could do the
job~\cite{DKS}. Here we give another form which can accelerate the
universe\footnote{This model was developed jointly with
J. Weller\cite{W}}. 
\begin{equation}
	V = \left[\frac{C}{(\phi - B)^2 + A}+ D\right]e^{-\lambda \phi}
	\label{eq:Branepotential}
\end{equation}
Such a potential could arise from the various brane models as a
Yukawa-like interaction  between branes\cite{DT} (we will call it throughout this paper the ``Brane'' model).

The right panel of Fig. \ref{twopotentials} shows the evolution of
densities with the Brane potential.
 During radiation era it behaves
like the AS potential. During the matter era however it retains significant density unlike
the AS potential. This has a significant impact on both structure formation (which is suppressed) and
the CMB as we shall see later. 

\begin{figure}[h]
\centerline{\hbox{\psfig{file=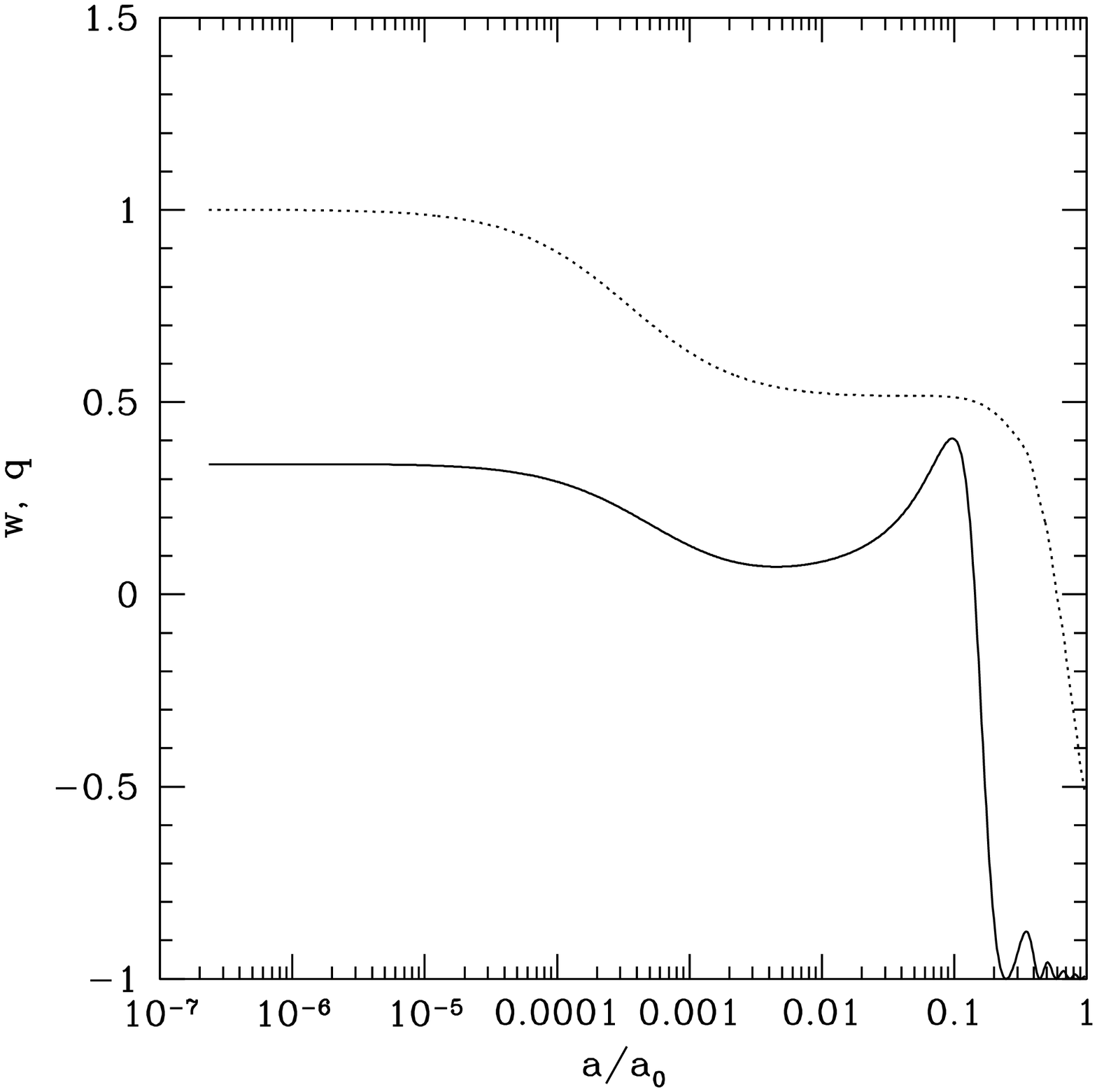,width=3.5in}\hspace{1cm}
\psfig{file=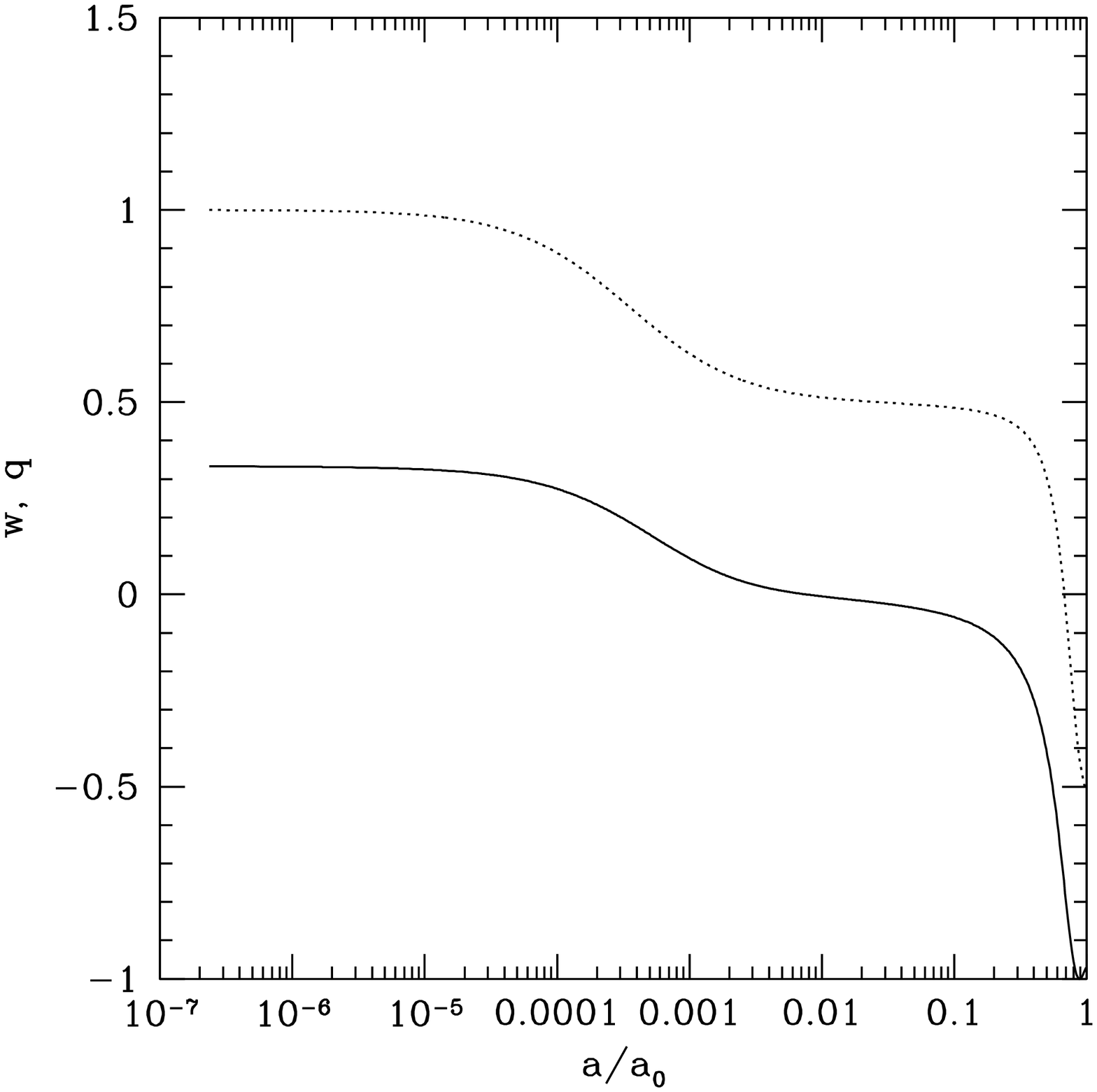,width=3.5in}}}
\caption{The equation of state parameter $w$ (solid) and deceleration
parameter  $q$ (dotted) for the same two models of
Fig. \ref{twopotentials}.  On the left we have the AS model and on the
right the Brane model. The difference of the two models seen in
Fig. \ref{twopotentials} is seen again here in $w$.(see the text for
more details)  Acceleration begins around $a = 0.7 - 0.9$ when $q$
becomes negative.} 
\label{twow}
\end{figure}

To understand why there is such a difference  we plot the equation of state parameter $w$ with
the scale factor in Fig. \ref{twow}. In the case of the AS model, soon after the field enters the 
matter era attractor where $w$ should go eventually to 0, the field
instead rushes toward the minimum of 
the potential and then undergoes damped oscillation. During that time 
the energy density of the field falls of significantly faster than the 
than energy density of the pressureless matter.
This behavior corresponds with a brief {\em rise} in $w(a)$. Due to
damping from the expansion however, the oscillations eventually stop, the field settles in
the minimum and  starts behaving like a cosmological constant. This is shown by the subsequent
oscillations of $w$ with decreasing amplitude until $w$ goes to $-1$. 

In the case of the Brane model the potential has a much sharper
minimum. In that case the field stays longer in the attractor regime
until suddenly it gets trapped in the minimum without oscillating. The
effect is that $w$ remains zero during most of the period of structure
formation (with the field in the matter attractor regime) and then
goes to $-1$ directly to start accelerating the universe. This allows
the field to retain the significant amount of energy density seen
previously. Moreover because of this, the Brane model starts
accelerating the  universe later than the AS model.

Another question concerning both potentials above is the stability of
the local minimum under quantum tunneling. It has been shown however
that tunneling is negligible which renders the minimum in both
potentials effectively stable~\cite{W}. 

Finally we should say that the tightest constraint comes from
requiring that $\Omega_\phi$ not be too large during
nucleosynthesis\cite{OSW} (at $a \approx 10^{-10}$). For
$\Omega_\phi(1MeV) \leq 0.1$ we get that $\lambda \geq 6.3$. The
models shown in Fig. \ref{twopotentials} do not obey this constraint
because they were chosen to dramatize the different behaviors.  When
comparing with data, we use more realistic parameters.  

\section{Density perturbations}
To make connection with the real universe one has to consider the growth of small
perturbations about an FRW metric. For extensive reviews see for example~\cite{Bar,KodSas,Mukhetal,MB}.
 Here we adopt the conventions of Ma and Bertschinger~\cite{MB}. We start with a 
line element given by
\begin{equation}
	ds^2 = a^2(\tau)[-d\tau^2 + (\delta_{ij} + h_{ij})dx^idx^j]
	\label{eq:metric}
\end{equation}
The metric perturbation is split into scalar, vector and tensor modes as usual.

\subsection{Perturbations in the scalar field}
 Here we give the relevant equations for the perturbations in the quintessence.
For an extensive review of scalar field perturbations in cosmology see for example~\cite{KodSas,Mukhetal}
and for the first applications in the case of quintessence~\cite{CDF,CDS}.
 
To include scalar fields into
the perturbations lets write the field as 
$\phi(\underline{x}) = \phi_0(\tau) + \phi_1(\underline{x})$ with $\phi_1<<\phi_0$,
where $\phi_0(\tau)$ represents the zeroth order homogeneous field i.e. $\nabla \phi_0 = 0$.
We can then form the stress-energy tensor from which we can read the
perturbed density $\rho_1$, pressure 
$P_1$ and velocity divergence $\theta$ of the field. 
It is important to note that the shear is zero independently of the form of the potential.
 The quantities of interest in $k-space$ are
\begin{eqnarray}
	\rho_1 &=& \frac{1}{a^2}\dot{\phi_0}\dot{\phi_1} + V_{,\phi} \phi_1 \\
	P_1  &=& \frac{1}{a^2}\dot{\phi_0}\dot{\phi_1} - V_{,\phi} \phi_1 \\
	\theta &=& \frac{k^2\phi_1}{\dot{\phi_0}}
\end{eqnarray}
The field perturbation also obeys a Klein-Gordon equation 
\begin{equation}
	\ddot{\phi_1} + 2\frac{\dot{a}}{a}\dot{\phi_1} + 
	(k^2 + a^2V_{,\phi\phi})\phi_1 + \frac{1}{2}\dot{h}\dot{\phi_0} = 0
	\label{eq:field}
\end{equation}
The above $2$nd order equation is equivalent to the two $1$st order equations (29) of ~\cite{MB}. One has
to keep in mind however that in general(as noted also in ~\cite{BMR}), the density perturbations 
in the field are not adiabatic i.e. $\frac{P_1}{\rho_1}$ is not in general equal to $\frac{dP}{d\rho}$.
We shall use the symbols $c_w$ and $c_s$ for $c_w^2 =\frac{P_1}{\rho_1}$ and $c_s^2=\frac{dP}{d\rho}$,
the later being the adiabatic speed of sound. If the field is adiabatic then the two quantities 
become equal $c_w = c_s$. 
The above point is important for understanding why even if quintessence has the same
equation of state as CDM it nevertheless does not clump. It all boils down to how the
Jeans length is defined. In the case of CDM, the Jeans length is zero by definition, contrary
to quintessence where it can be as large as the horizon. This is
because the Jeans length must be defined in terms of $c_w$ and not $c_s$ to include
the effect of entropy perturbations. In the case of quintessence it is exactly this
effect which prevents clumping. Basically, the quintessence field is
able to have a non-zero pressure perturbations even when $c_s=0$, and
the pressure will resist gravitational collapse.

Another thing to note about the field equation (\ref{eq:field}) is the 
physical meaning of the various terms. 
The Hubble expansion provides a time-dependent ``drag'' term and the potential a 
time dependent ``mass'' term. The last term gives the gravitational
sources for the field perturbations.

\subsection{Growing super-horizon perturbations in the radiation era}
In our model, deep in the radiation era we can assume that the field is in the attractor
 and that the potential is a pure exponential. In this case we can use the analysis of 
Ferreira and Joyce~\cite{FJ} to get the initial conditions. We should note that the initial 
conditions given here are curvature initial conditions. 

We can use (\ref{eq:att_quant}) with $n=4$ for the radiation era. Then we have
 for Quintessence
\begin{eqnarray}
	\rho_\phi &=& 3V = \frac{3}{4}\dot{\phi_0}^2 \\
	\frac{\dot{a}}{a} &=& \frac{1}{\tau} \\
	\Omega_\phi &=& \frac{4}{\lambda^2} \\ 
	\delta_\phi &=& \frac{\lambda}{3}(\tau\dot{\phi_1} - \phi_1) \\
	c_{w\phi}^2 &=& \frac{\tau\dot{\phi_1} + \phi_1}{\tau\dot{\phi_1} - \phi_1} \\
	\tau^2\ddot{\phi_1} + 2\tau\dot{\phi_1} &+&
		 [ (k\tau)^2 + 4]\phi_1 + \frac{2}{\lambda}\tau\dot{h} =0 
\end{eqnarray}

Assuming that only photons and quintessence are dominant during that period we find 
for the gravitational perturbation $h$ 
\begin{equation}
	\tau^2\ddot{h} + \tau\dot{h} + 6\Omega_\gamma\delta_\gamma + 
       3(1 + 3c_{w\phi}^2)\Omega_\phi\delta_\phi = 0
\end{equation}
For superhorison perturbations we assume $k\tau<<1$ and therefore to 
first order $\theta_\gamma = 0$.Therefore
\[ \delta_\gamma = -\frac{2}{3}h\]

Combining the last six equations (for $k^2=0$) we get
\begin{eqnarray}
	\tau^2\ddot{h} + \tau\dot{h} -4(1 - \frac{4}{\lambda^2})h 
		+ \frac{8}{\lambda}(2\tau\dot{\phi_1} + \phi_1) &=& 0 \\
	\tau^2\ddot{\phi_1} + 2\tau\dot{\phi_1} + 4\phi_1 + \frac{2}{\lambda}\tau\dot{h}&=& 0  
\end{eqnarray}
Let $h = C\tau^m$ and $\phi_1 = D \tau^m$.
Then


\begin{equation}
	m = \pm 2 , \qquad m = \frac{-1 \pm \sqrt{\frac{64}{\lambda^2} - 15}}{2} 
\end{equation}

The growing physical mode is the one proportional to $\tau^2$.
For the coefficients $C$ and $D$ we get (a result also found by Ferreira and Joyce)
\begin{equation}
	D = -\frac{2}{5\lambda} C
\end{equation}

This gives
\begin{equation}
	\delta_\phi = \frac{4}{15}\delta_c
	\label{eq:delta_ratio}
\end{equation}

Note that the above result gives $c_w^2 = 3 \neq c_s^2= w = \frac{1}{3}$ so quintessence 
is not adiabatic even if $\dot{w}=0$. 

\subsection{Initial Conditions}
From the above analysis we can form the initial conditions following
the procedure in Ma and Bertschinger. 
The initial conditions for all the standard quantities remain the same. The two additional initial
conditions for quintessence are

\begin{equation}
	\phi_1 = -\frac{2}{5\lambda}h, \qquad \dot{\phi_1} = -\frac{2}{5\lambda}\dot{h}
\end{equation}

Even if we don't use the correct initial conditions for the field, as a property of the
growing curvature mode, the correct evolution of the field density contrast will be reached
quite rapidly and the initial conditions given above will be valid.

\section{Structure formation}
First we investigate the growth of structure with time using the density perturbations 
as a guide. Then based on that investigation we give the resulting matter power spectrum
and explain its form.
\subsection{The growth of structure} 
\begin{figure}[h]
\centerline{\hbox{\psfig{file=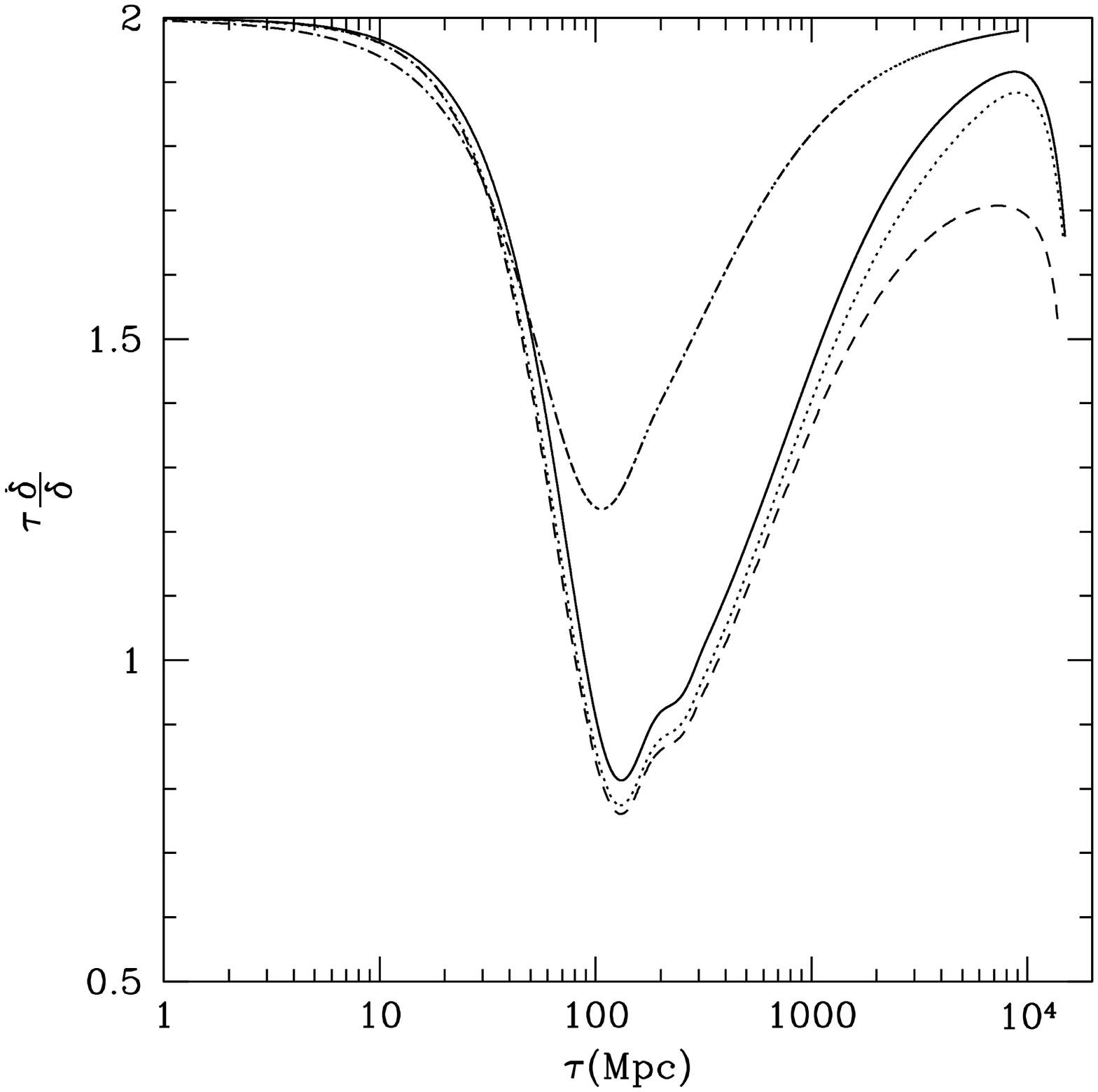,width=3.5in}\hspace{1cm}
\psfig{file=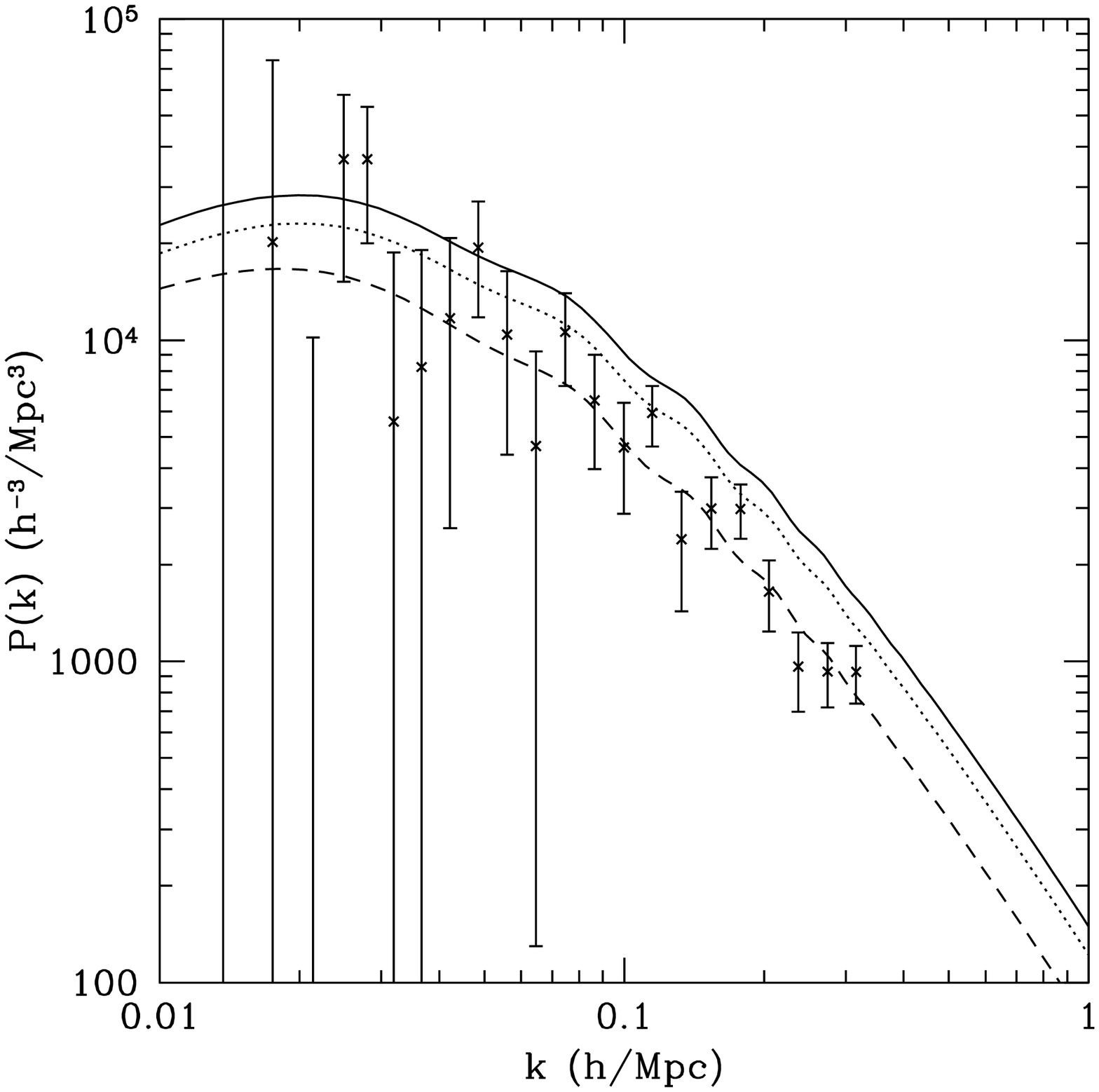,width=3.5in}}}
\caption{{\bf LEFT:} The effective dimensionless growth rate
$n_{\text{eff}}$ for four different models at $k=0.1Mpc^{-1}$. 
At early times $n_{\text{eff}}\rightarrow 2$ for all models, 
due to the $\tau^2$ dependence of the growing superhorizon modes. Growth is
suppressed in the radiation era ($\tau \approx 100 Mpc$) and in the
$\phi$-era (today). In the 
matter era sCDM(dash-dotted) has the most growth, then $\Lambda$CDM
(solid), AS (dotted)
 and finally with the least growth the Brane (dashed).  The last two
are the same models shown in Fig. \ref{twopotentials}.\protect\newline
{\bf RIGHT:} The matter power spectrum normalized to COBE using the Bunn-White fitting formula.
 The plotted models are
$\Lambda$ (solid), AS (dotted) 
Brane (dashed). All the models have $h=0.75$, $\Omega_c = 0.297$, $\Omega_b = 0.053$ and 
$\Omega_{\phi}=0.65$. Here we use more realistic quintessence parameters consistent
nucleosynthesis. In  
particular for the AS model $\lambda=8$, $B=33.9627$, $A=0.01$ and for the Brane model
$\lambda=8$, $B=35.13689$, $C=0.01$ and $D=0.01$.
The data points are the decorrelated data of Hamilton et.al.~\protect\cite{HT,HTP}}
\label{growth}
\end{figure}

One intuitive and important quantity is the dimensionless growth rate for
the cold dark matter,
\begin{equation}
 n_{\text{eff}}=\tau\frac{\dot{\delta_c}}{\delta_c}.
 \label{eq:growth}
\end{equation}
This quantity give an instantaneous measure of the growth rate of
structure. The time evolution of $ n_{\text{eff}}$ is shown for a
number of models on the left panel of Fig. \ref{growth}.  
Initially, outside the horizon, $n_{\text{eff}}=2$ completely independent of the model of
structure formation(for a flat universe). This is due to the fact that because of 
causality the form of the background energy density is not relevant, only the amount.
After entering the horizon, the growth of structure is suppressed
during the radiation era simply because dark matter is extremely
sub-dominant. During the matter era, dark matter becomes 
dominant again and in the case of an SCDM model, $n_{\text{eff}}$
eventually reaches  the value of $2$ again. For a $\Lambda$CDM model
however, $n_{\text{eff}}$ after growing for a while, it eventually
drops down to zero. This happens when the universe enters the
accelerating era where the perturbations leave the horizon and
structure stops forming.

The two quintessence models considered here have a similar effect to a $\Lambda$CDM model. 
The only difference is the existence of a significant amount of dark energy during the matter
era. This results in a suppression of structure formation since the amount of dark matter
is less and therefore structure forms less efficiently. As stressed in
section III.A, even when $w_\phi=0$ quintessence can have pressure
perturbations which allow it to resist clumping.

 This is not very important for the AS potential since 
if you notice from Fig. \ref{twopotentials}, $\Omega_\phi$
goes very close to zero before the accelerating era. Therefore in that model structure is just
a little bit suppressed compared to a $\Lambda$CDM model, but not by much. 

For the Brane potential however the above effect is much stronger.
In fact $\Omega_\phi$ stays quite significant during all of the matter era. Therefore for that 
potential, structure growth is even more suppressed as one can see from Fig. \ref{growth}. Not 
only do we have the acceleration era like before which suppresses structure anyway, we also
have a significant amount of quintessence which doesn't cluster. This means that the amount of
clustering matter is even less in this case so structure is even more
suppressed. 

\subsection{The matter power spectrum}
Based on the reasoning of the previous section, we can predict the form of the matter 
power spectrum. For the AS model we expect the power to be a bit less
than for $\Lambda$ model with the same cosmological parameters. The
Brane model should exhibit additional suppression. This is indeed the
case as one can see in the right panel of Fig. \ref{growth}.

If we use a Hubble constant of $75 kms^{-1}Mpc^{-1}$, even though structure is suppressed for
the Brane model, it fits the data very nicely. All of the models shown in the figure have
$h=0.75$, $\Omega_c=0.297$, $\Omega_b = 0.053$, $\Omega_\phi = 0.65$ and some
reionization with $\tau_{\text{opt}} = 0.1$. The corresponding CMB anisotropies are shown later in the
right panel of Fig. \ref{cls}.
The $\sigma_8$'s  1.08,0.98 and 0.77  for $\Lambda$CDM,
 AS and Brane potentials respectively. 
\begin{figure}[h]
\centerline{\hbox{\psfig{file=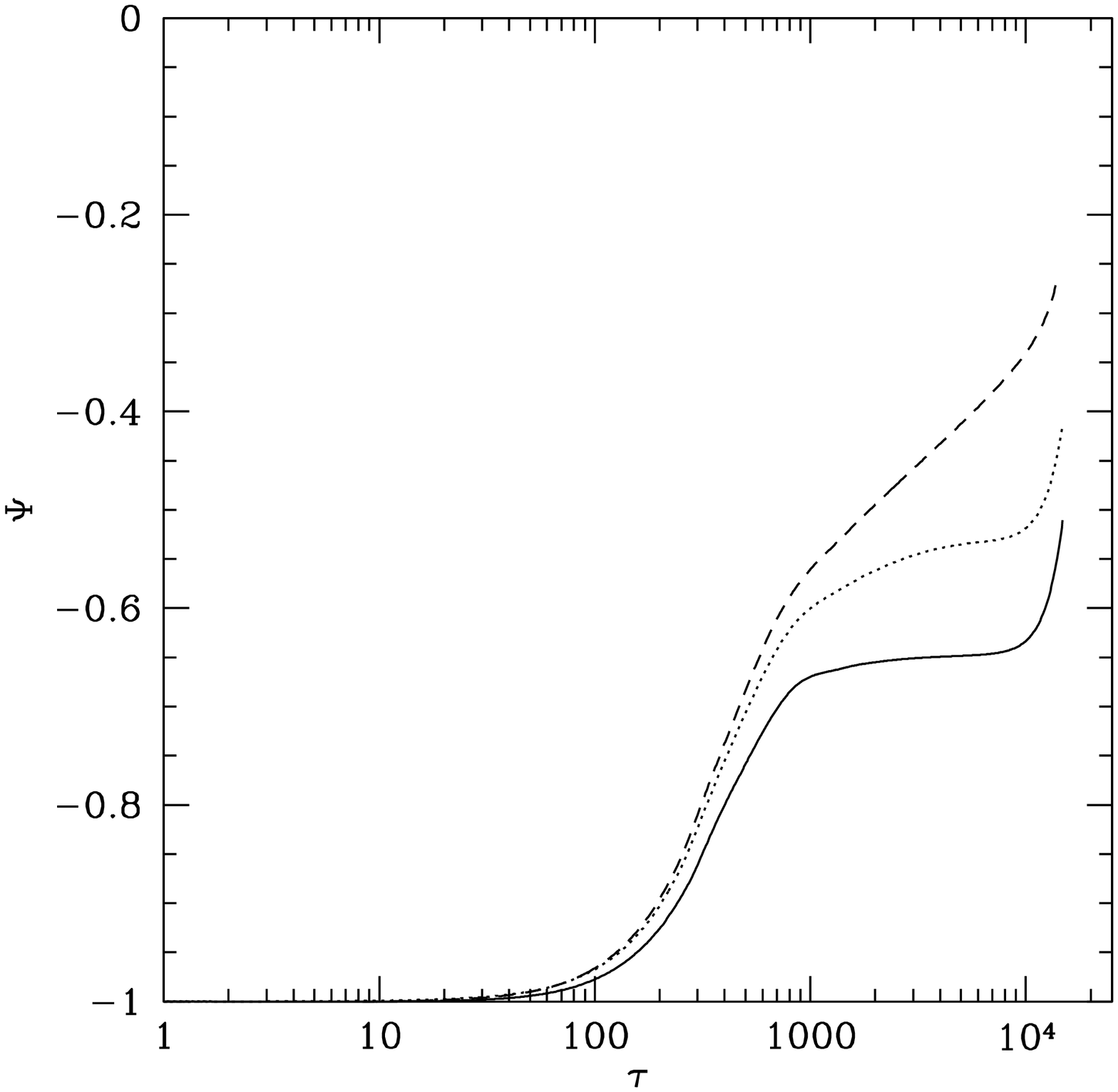,width=3.5in}\hspace{1cm}
\psfig{file=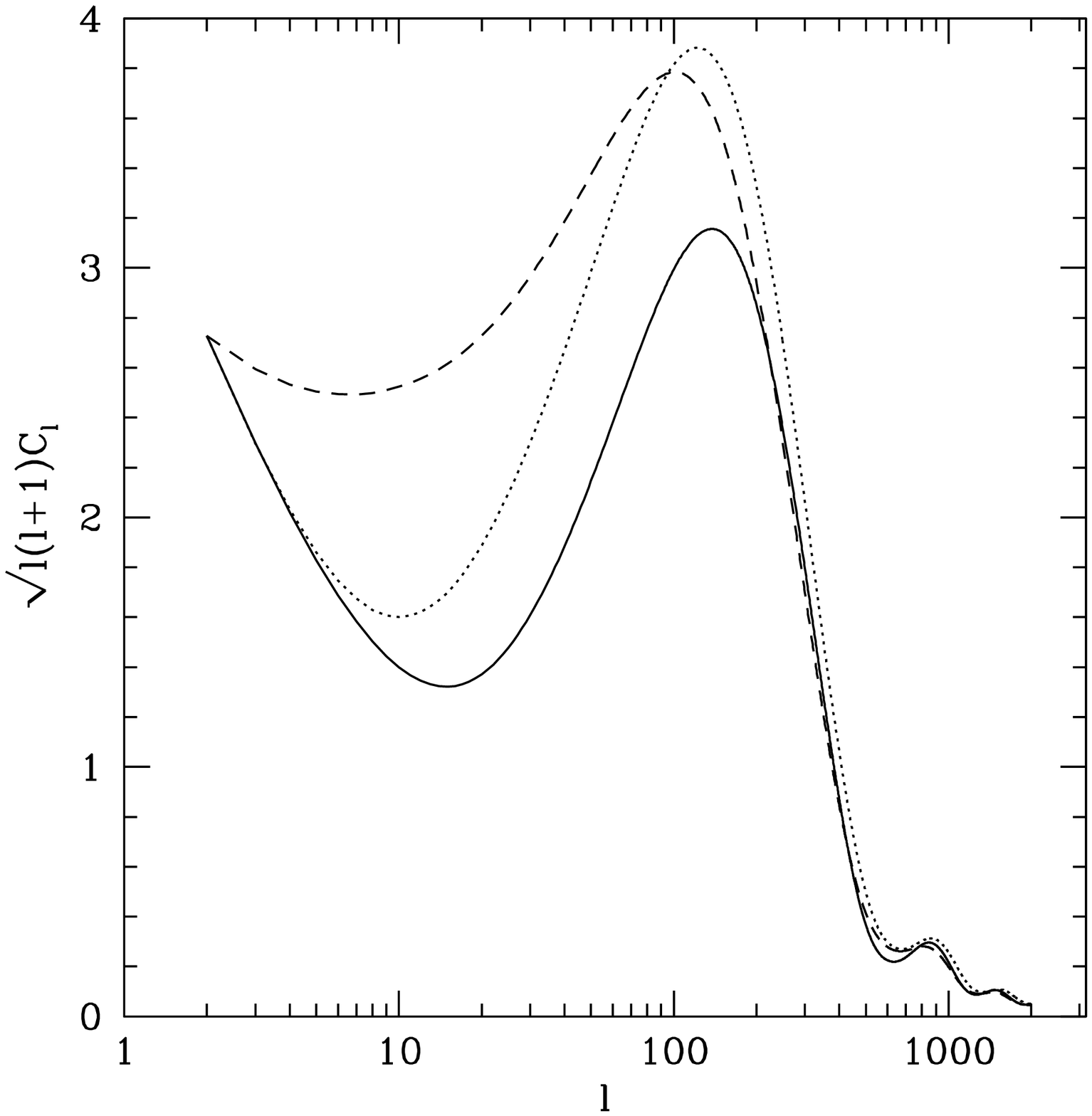,width=3.5in}}}
\caption{{\bf LEFT:} The Newtonian potential $\psi(\tau)$ at $k=0.01Mpc^{-1}$, normalized to $-1$ as 
$k\tau\rightarrow 0$. Shown are the same three models of
Fig. \ref{twopotentials}:  AS model (dotted), 
$\Lambda$ (solid), Brane model (dashed). Notice
the very strong decay of the potential in the case of the Brane model,
 which leads to a very strong ISW effect. \protect\newline
{\bf RIGHT:} An illustration of the ISW effect for the same models. All models have the same initial
power spectrum. The
strong ISW for the Brane model at COBE scales affects the normalization and reduces the final 
anisotropies.}
\label{newton}
\end{figure}

It is not the goal of this paper to scan all of parameter
space and compare a wide range of models, and we are not suggesting that
any of the models in Fig. \ref{growth} are the best fitting model 
of their type.  Figure \ref{growth} is intended to illustrate the
physical differences of the different types of models by holding as
many aspects as possible constant across the four models.  Figure
\ref{growth} also illustrates the somewhat surprising result that it
is possible for models with significant contributions from the dark
energy at early times (i.e.the Brane model) to provide a good fit to the data.

\section{The Cosmic Microwave Background}
Lastly we give the CMB anisotropy power spectrum and investigate the
various physical effects due to the field. We follow the analysis of
Hu and Sugiyama ~\cite{HS1,HS2,HS3}.  
Following the common practice we divide the CMB anisotropies into primary and secondary.
The primary anisotropies are the ones formed at the Last Scattering
Surface (LSS), the
secondary being the ones due to the subsequent cosmological evolution. Due to significant
quintessence energy density at the LSS, the primary anisotropies are different compared with
other models. In this section we drop sCDM from consideration and
compare the remaining three models discussed above: the AS and Brane
models, and the $\Lambda$CDM model with similar parameters. Moreover, since we are only interested
in the differences cause by quintessence, we keep other cosmological
parameters($\Omega_b$, $\Omega_c$ and $H_0$) fixed. 

\subsection{Primary anisotropies}
With the above parameters fixed, the only other parameter that can affect the primary anisotropies
is the Hubble parameter during the time of LSS and earlier. Changing the Hubble parameter affects
the heights of the peaks through three main effects, the driving effect, the doppler shift and
Silk damping.

Looking at fixed $a$ it is easy to show that the Hubble parameter in a quintessence model $H_\phi(a)$
is related to the Hubble parameter in a $\Lambda$ model $H_\Lambda(a)$ at the same $a$, by
\begin{equation}
	H_\phi = H_\Lambda\sqrt{\frac{1 - \Omega_\Lambda}{1 - \Omega_\phi}}
	\label{eq:Phi_Lambda_relation}
\end{equation}
 Therefore, since for the quintessence models under consideration here
$\Omega_\phi$ is quite  
 significant during all the history of the universe(where as  $\Omega_\Lambda$ isn't),
 the Hubble parameter in a quintessence
universe will in general be larger than the one in a $\Lambda$ universe. The above statement is
of course true only if we keep $\Omega_{0m}$, $\Omega_{0\phi}$ and $H_0$ fixed which is what we
do for comparing the models. 
\begin{figure}[h]
\centerline{\hbox{\psfig{file=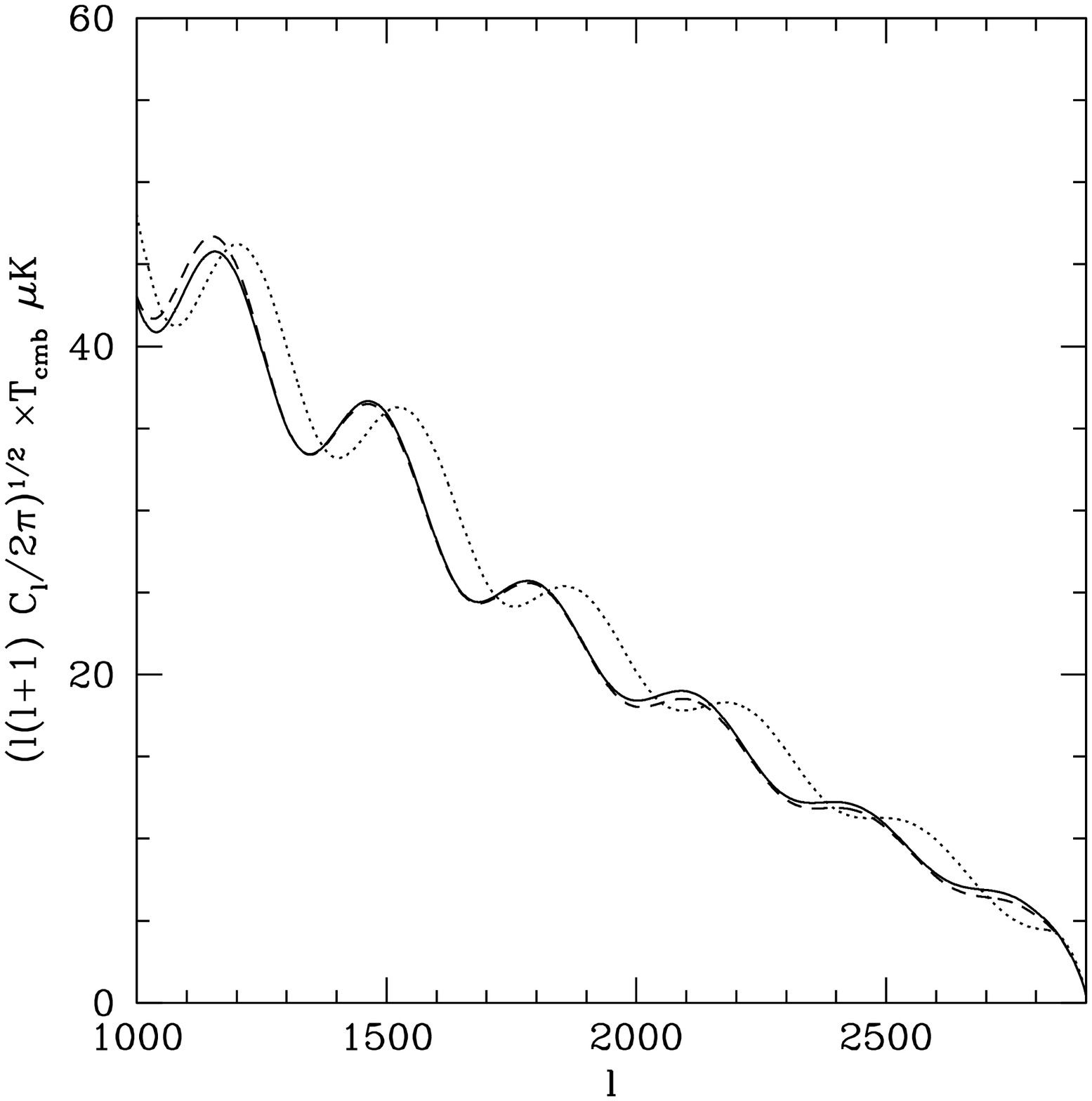,width=3.5in}\hspace{1cm} 
\psfig{file=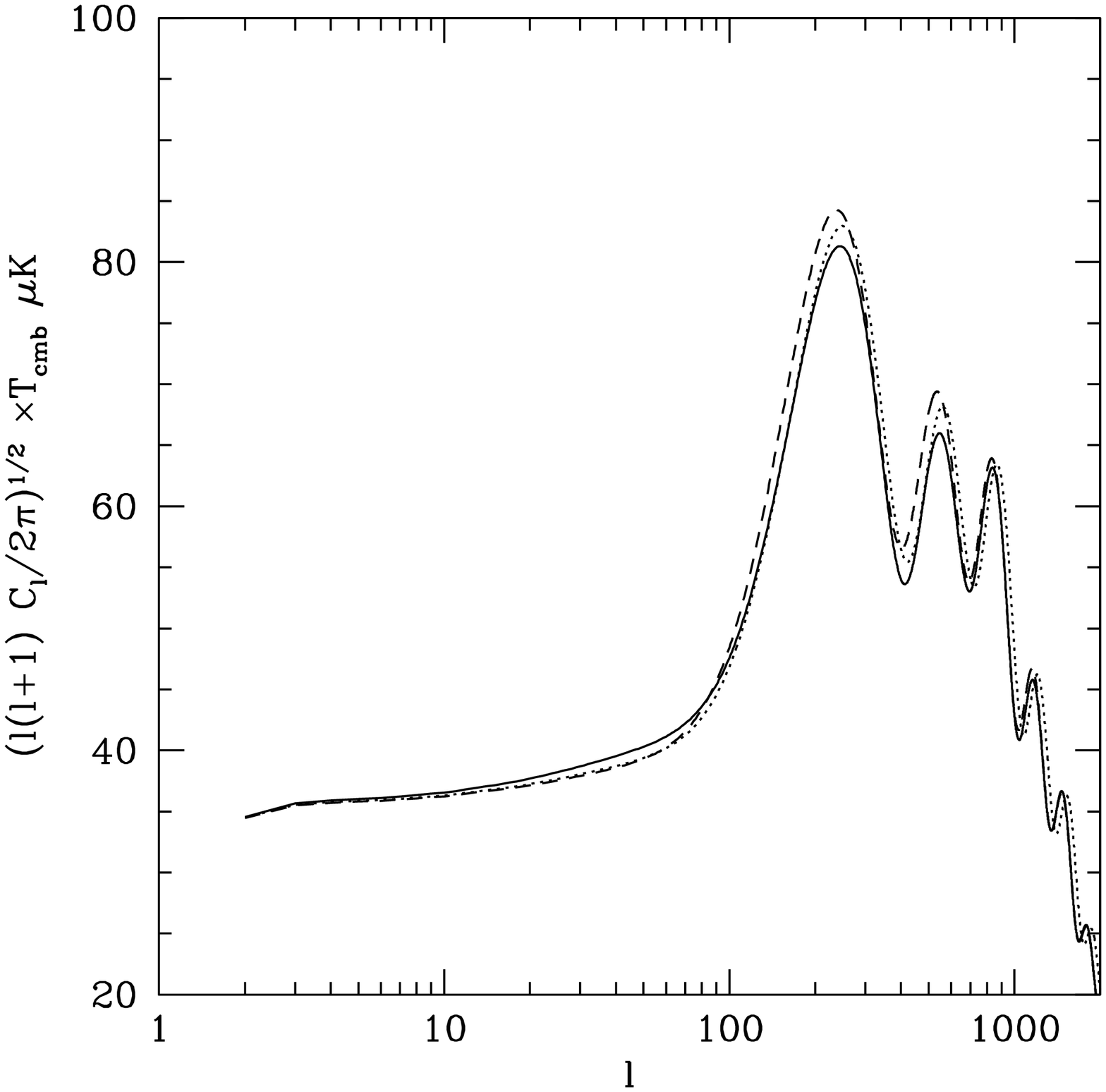,width=3.5in}}}
\caption{{\bf LEFT:} The effect of Silk damping on the total primary anisotropies for the same models
as Fig.\protect\ref{twopotentials} with the same initial power spectrum.
 In the case of quintessence, Silk damping is slightly more effective(see text).
{\bf RIGHT:} Primary temperature anisotropies for the same models, as would have been
observed today if there was no other anisotropy source. The driving effect is observable on the first
four peaks. The much higher 2nd peak for quintessence as well as the trough before it shows the Doppler 
effect.}
\label{Silk}
\end{figure}

 The driving effect arises from a decaying Newtonian potential $\Psi$ before the LSS (see
Fig. \ref{newton}), as well as a dilation effect from a decaying space curvature $\Phi$(not to be
confused with $\phi$ which is the quintessence field).
The Newtonian potential and the curvature can be written in terms of the other perturbations as
\begin{eqnarray}
	\Phi &=&  -\frac{3a^2H^2}{2k^2} 
               \sum_i[\Omega_i\delta_i + \frac{3aH}{k^2}(1+w_i)\Omega_i\theta_i] \nonumber \\
	\Psi &=& \Phi - \frac{6a^2H^2}{k^2} (\Omega_\gamma\sigma_\gamma + 3\Omega_\nu\sigma\nu)
   \label{eq:Newtonian}
\end{eqnarray}
The effect of quintessence on the driving effect comes from the significant fractional
densities of quintessence during the radiation and matter era. During the radiation era
$\Omega_\phi$ is always greater than the on during matter era. This results in a greater 
difference of $H_\phi$ from $H_\Lambda$ in the radiation era than in the matter era. 
The result is both the Newtonian potential and curvature decay faster in the case of 
quintessence. The effect is an increased
driving effect in the case of quintessence which increases the temperature anisotropies for
all peaks. 
 We should also note that the same effect occurs by having more species of relativistic
 particles around(e.g. more massless neutrinos), the reason being exactly the same.
 Moreover this is more significant for the Brane potential whose Hubble parameter is always
greater that the one for the AS potential. 

The velocity perturbation makes a contribution to the temperature
anisotropies that is $90$
degrees out of phase with the contribution from the density
perturbation. The velocity is 
sub-dominant compared to the density which makes the effect
weak. There is however a small but observable difference between a
$\Lambda$ model and quintessence. For quintessence even peaks as well
as the troughs before them are slightly raised compared to a $\Lambda$
model. This causes the oscillations appear to be ``fattened'' up  due to
the modulation of the velocity with the density to get the final
temperature anisotropy.

Finally we have Silk damping which is different in the three models. The damping coefficient
 $k_D$ is given by
\begin{equation}
	k_D^{-2}(a) = \frac{1}{6}\int_0^a \frac{1}{a^4H^2\tau'}
		\frac{R^2+\frac{4}{5}(1+R)}{(1+R)^2} da 
	\label{eq:Silk_damping}
\end{equation}
where $R= \frac{3\rho_b}{4\rho_\gamma}$ and $\tau_{\text{opt}}$ 
is the optical depth. The damping coefficient
is proportional to the Hubble parameter and is therefore smaller in the case of
 quintessence. This means that damping becomes more effective at smaller $k$(larger scales)
in the quintessence models. The effect will be to suppress the small scale temperature anisotropies 
more effectively.
\begin{figure}[h]
\centerline{\hbox{\psfig{file=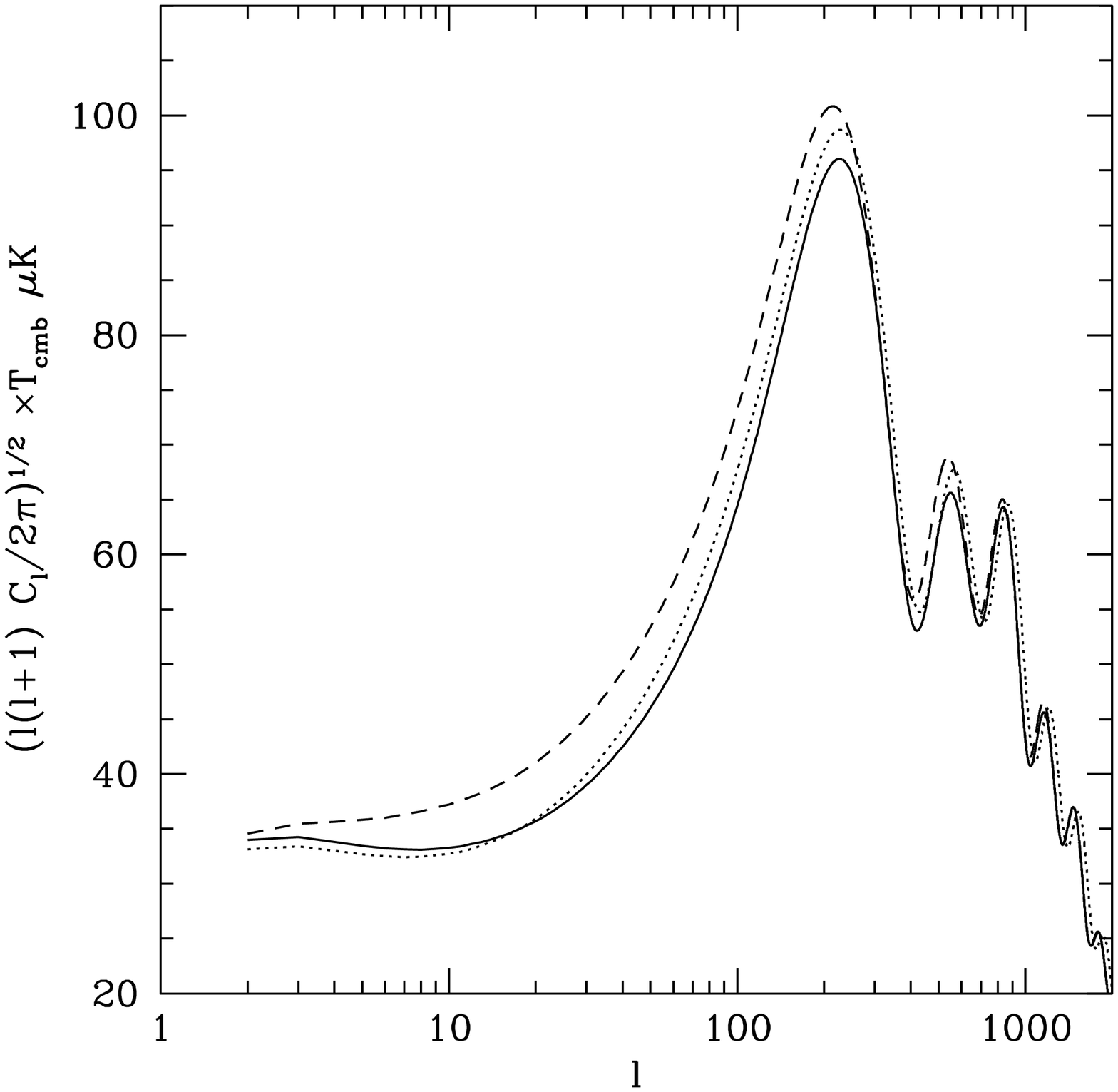,width=3.5in}\hspace{1cm}\psfig{file=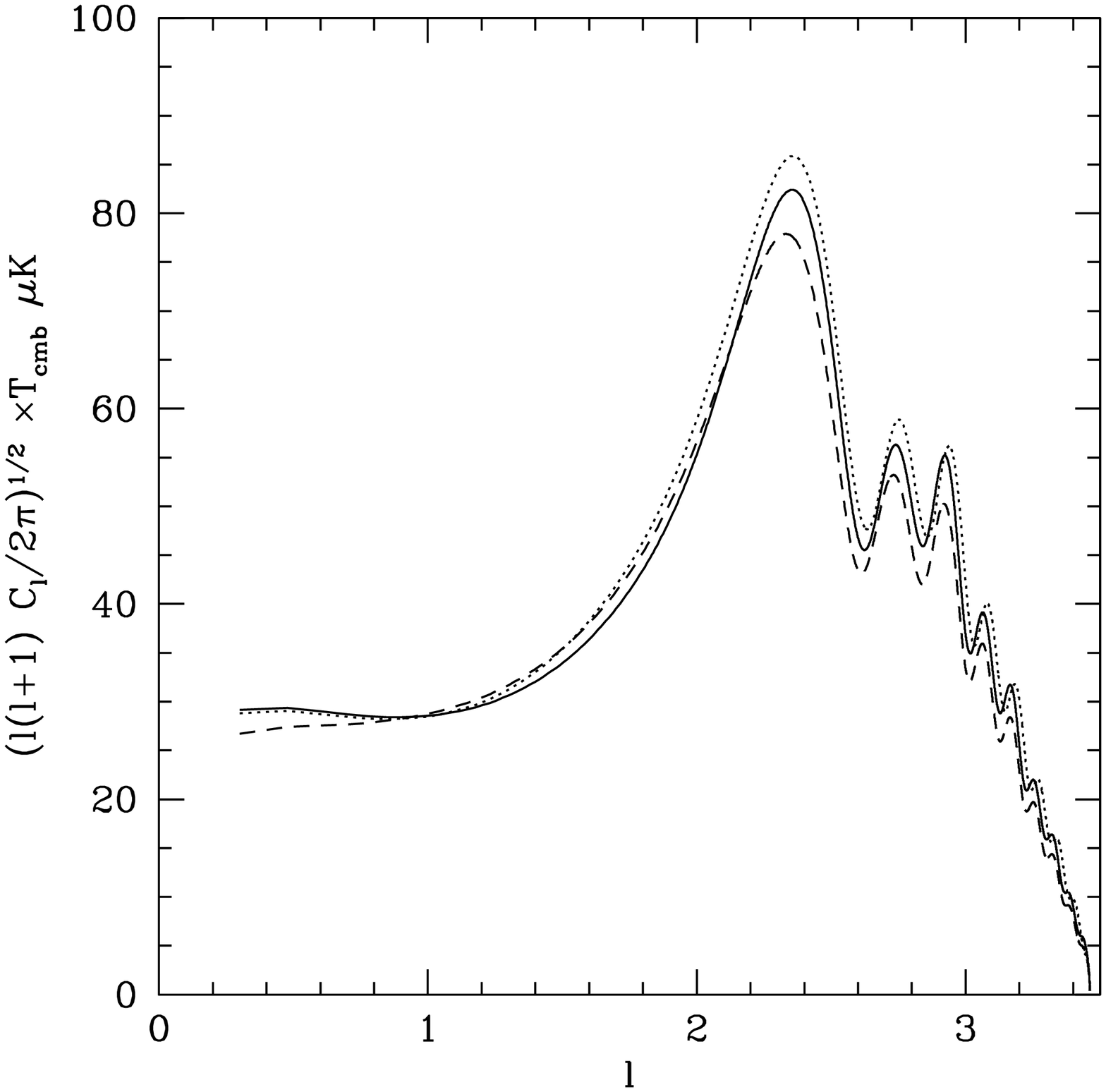,width=3.5in}}} 
\caption{The CMB anisotropies for the models of Fig.\protect\ref{twopotentials}. Shown on the left are
the anisotropies with the same initial power spectrum and on the right after COBE normalizing. Due to
the strong ISW the anisotropies of the Brane model (dashed) are suppressed after normalization.}
\label{cls}
\end{figure}
Since at large $k$ the driving effect eventually becomes unimportant,
 Silk damping eventually becomes stronger than the driving effect at small 
scales. This makes a $\Lambda$ model with equivalent cosmological parameters
have larger anisotropies than  quintessence even though  
quintessence has larger anisotropies in the first two peaks.  The total primary 
anisotropies are shown in Fig.\ref{Silk}

Let us now consider the secondary anisotropies which can change this
picture quite a lot. 

\subsection{Secondary anisotropies}
The most notable secondary anisotropy is the Integrated Sachs-Wolfe
(ISW) effect. To dramatize this effect for pedagogical purposes  we
first consider the models from Fig. \ref{twopotentials} which have
a higher proportion of Quintessence than is allowed by nucleosynthesis 
(and an equally bad $\Lambda$ model).

The ISW comes from a decaying Newtonian potential after the LSS (see Fig. \ref{newton}).  
The standard treatment breaks the ISW down to early and late ISW. In
our case however we have an intermediate ISW as well.

 The right panel of Fig. \ref{newton} shows ISW contribution  for the three models while the
left panel of Fig. \ref{cls} shows the angular power spectrum for  the
same models.  The expression for $\Theta^{isw}$ is the ISW contribution
to the total anisotropies as defined in Hu and Sugiyama~\cite{HS2} eq.11. 
  
The ISW effect is very important for the Brane model (shown in all the relevant figures 
with a dashed curve). If what we saw today was just primary anisotropies 
then this model should have had the highest first peak of all three models. What we see
 however is exactly the opposite: it has the lowest peak. 
This is completely due to a very strong ISW which is shown on the right panel 
of Fig. \ref{newton}. 
The ISW for the Brane model boosts power at large scales($l<100$) compared to the other models.
Therefore it will have a direct effect on COBE normalization(or in fact any other kind of 
normalization which includes points at those scales)- the anisotropies
 have to be scaled down to fit COBE. What happens physically is that to get the observed 
power, one has to use a smaller amplitude for the initial power spectrum. This makes the 
 the ratio of the first peak to the
spectrum at COBE scales smaller in the case of the Brane model.
 Normalizing to COBE then suppresses the small scale anisotropies.
So even though the model had the highest peak at the LSS, it
now has the smallest. 

The second effect of the very strong ISW on the Brane model is to shift the first acoustic peak 
to larger scales. This is though of lesser magnitude than the same effect caused by the
smaller angular diameter distance.
\begin{figure}[h]
\centerline{\hbox{\psfig{file=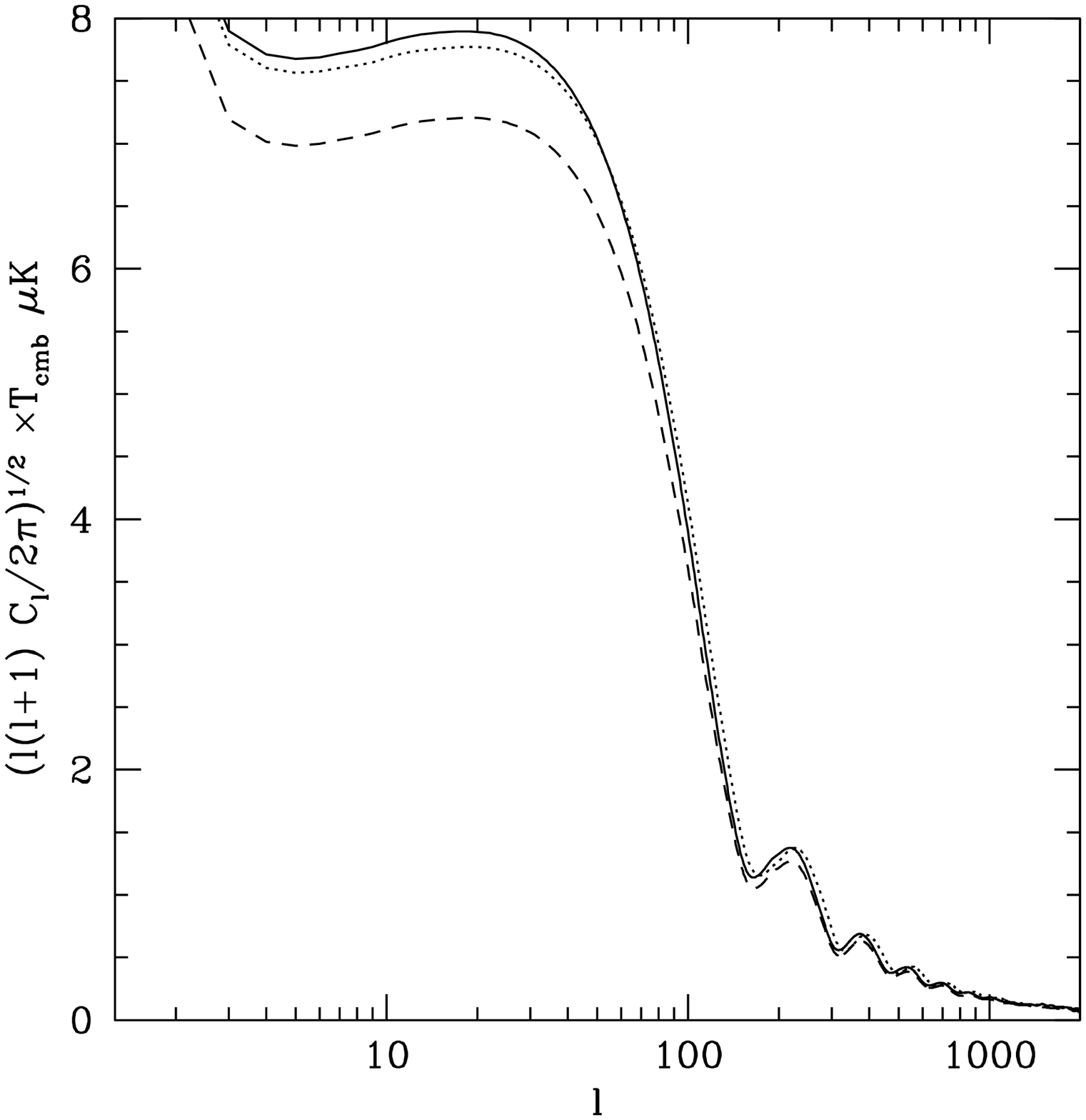,width=3.5in}\hspace{1cm}
\psfig{file=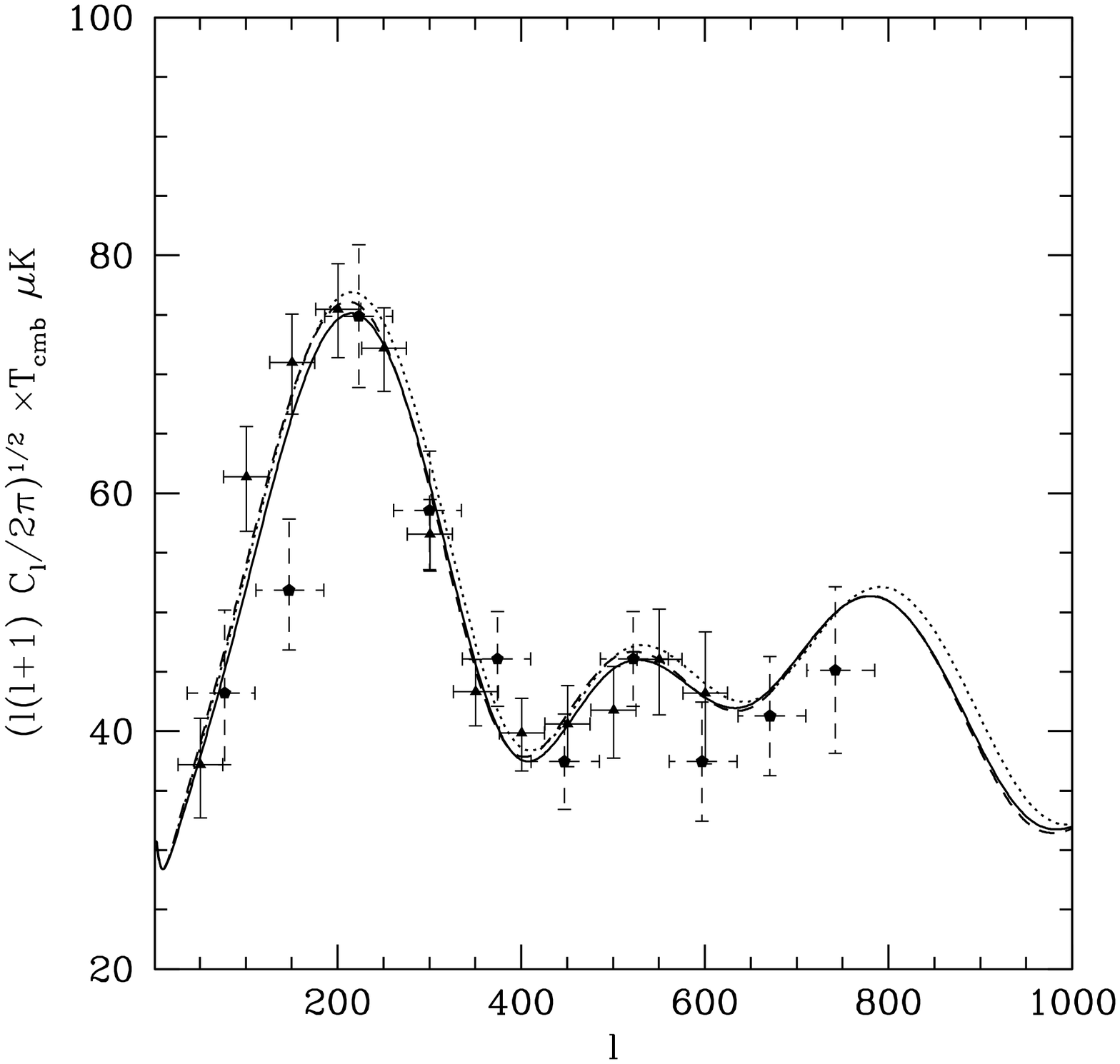,width=3.5in}}}
\caption{{\bf LEFT:} Tensor $C_l$s for cosmological constant and the two quintessence
models. All models have $h=0.65$, $\Omega_b=0.053$ $\Omega_c=0.247$, $\Omega_\phi=0.7$, $n_s=0.98$,
$n_t=-0.02$ and ratio of tensor to scalar quadrupole of $0.1$.
Shown is a $\Lambda$CDM(solid), AS (dotted) with $\lambda=5$, $B=54.4057$, $A = .01$
and Brane (dashed) model with $\lambda=5$, $B=56.10425$, $A = .01$, $C=1$ and $D=0.1$.\protect\newline
{\bf RIGHT:} The CMB anisotropies for the more realistic models of Fig. \ref{growth}(Right panel). The
effects discussed in the text are still visible but not as strong. Shown are the 
BOOMERanG98(solid)~\protect\cite{BOOMERanG} and MAXIMA(dashed)~\protect\cite{MAXIMA} data.}
\label{tensor}
\end{figure}

Finally we also have gravitational lensing on the CMB photons. This is more 
significant on smaller scales. In the case of polarization the effect is smaller than
$10\%$ and for the temperature anisotropies even smaller~\cite{ZS}. 
The effect is a smearing of the 
peaks due to mixing of $l$-values without destroying the overall
structure of the peaks. We calculate it using the linear evolution
method of CMBFAST and assuming that quintessence will have the same
effect on lensing as a pure cosmological constant. This is reasonable
since like a cosmological constant, quintessence doesn't cluster and
therefore its power spectrum does not contribute to lensing. The
quintessence power is actually much smaller than the photon or neutrino power
which themselves make a vary small contribution and are neglected. 

\subsection{Position of the peaks}
 The position of the peaks depends on two quantities, the sound horizon and the angular diameter distance
to LSS. Both are affected by the presence of quintessence. 

First lets consider the sound horizon. Even though the baryon density is kept the same,
 hence the speed of sound is the same in all three models, the time to the
 LSS is different due to the different expansion rate. Hence the sound horizon given by
\begin{equation}
	r_{s} = \int_0^{\tau_{*}} c_s d\tau
	\label{eq:SoundHor}
\end{equation}
is different($\tau_{*}$ is the time at last scattering).
 More specifically it is inversely proportional to $H$ which means that quintessence
has a smaller sound horizon. This shifts the anisotropies at LSS(in $k$-space) to smaller scales.
Greater $H$ also makes the time at LSS $\tau_{*}$ smaller.
\begin{figure}[h]
\centerline{\hbox{\psfig{file=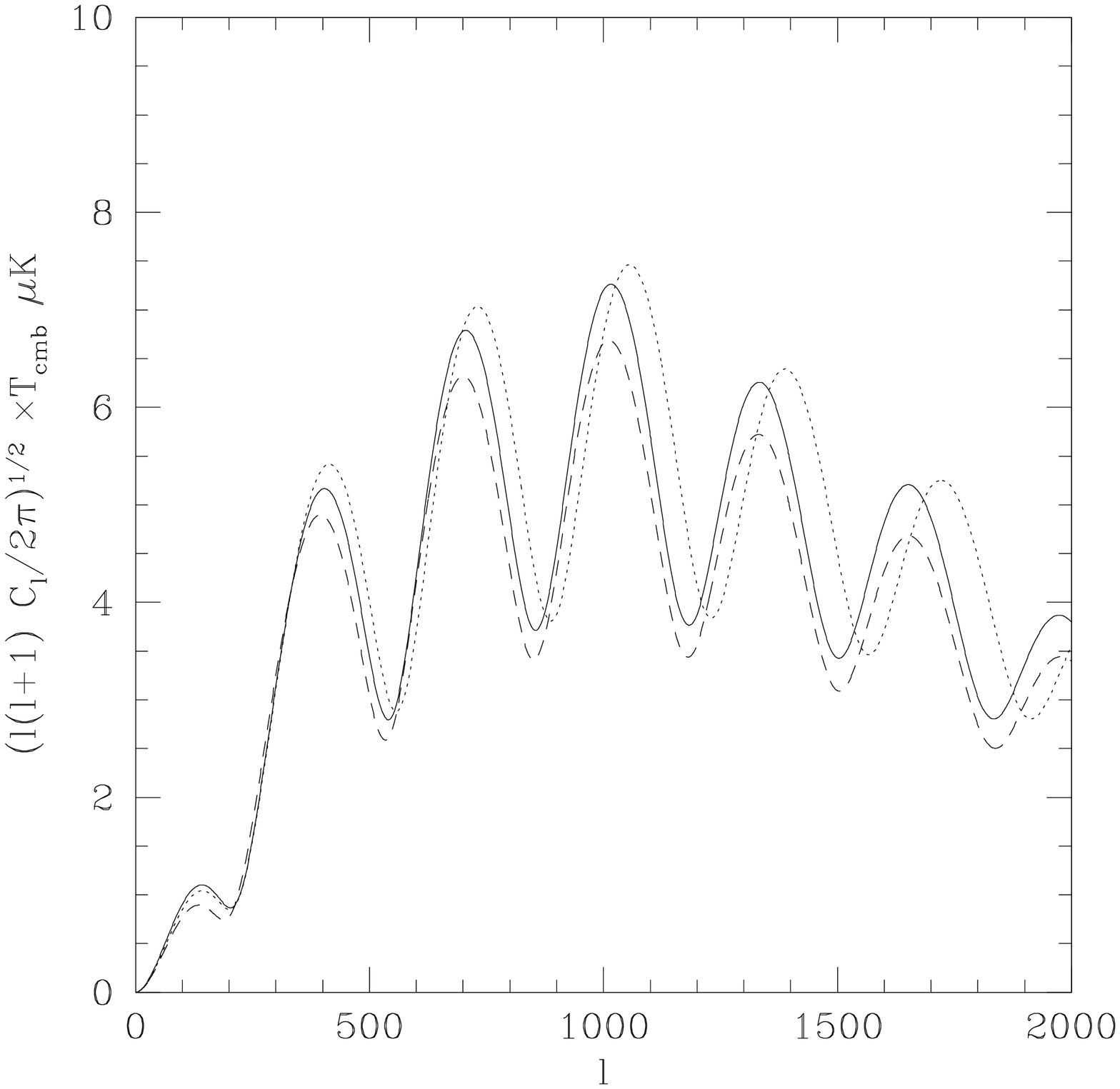,width=3.5in}\hspace{1cm}
\psfig{file=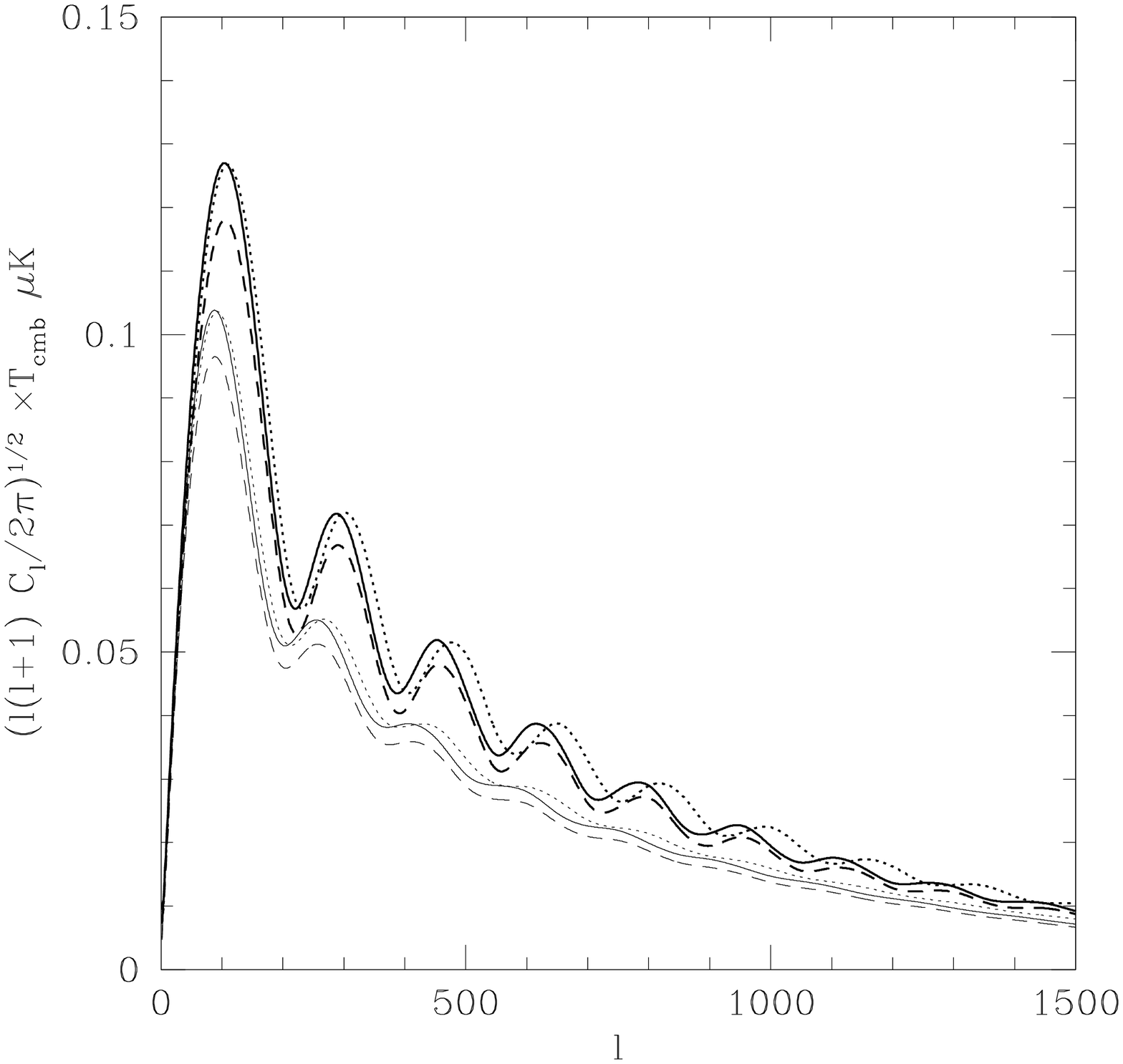,width=3.5in}}}
\caption{{\bf LEFT:} E-type polarization for the same models as Fig. \ref{twopotentials}. The strong ISW in
 the case of the Brane model suppresses indirectly the polarization spectrum. \protect\newline
{\bf RIGHT:} Tensor E-type(heavy curves) and B-type(light curves) polarization for the models same models 
as Fig.\protect\ref{tensor}(left panel). 
Since there is no ISW the difference in the AS and $\Lambda$ models is just in the frequency of oscillation. 
For the Brane model we do have however the indirect effect of a strong ISW through normalization of the
scalar $C_l$s amplitude}
\label{polar}
\end{figure}

Next we have the angular diameter distance to LSS which is smaller in the case of
quintessence. For the AS model it is not much smaller than a $\Lambda$ model since the universe
starts accelerating at around the same time in both models. In the case of the Brane model however
that universe is younger and therefore the angular diameter distance to LSS, smaller. In fact
the more quintessence we have during the history of the universe, the younger it is since the
universe starts accelerating later. If we have enough quintessence it doesn't accelerate at all.
This appears to be counterintuitive. Remember however that in order to have more quintessence
around one has to make the potential shallower. If it is shallow enough the field just rolls 
through.

From the analysis of ~\cite{HS2} the difference in $l$ between peaks is given by
\begin{equation}
	\delta l = \pi \frac{\tau_{0} - \tau_{*}}{r_s}
	\label{eq:deltapeak}
\end{equation}
where $\tau_0$ is the time today.
For the models of the right panel of Fig. \ref{cls} 
we get 292, 297 and 293 which is in very good agreement with 
the actual numerical result. The AS model has roughly the same angular
diameter distance as the $\Lambda$CDM model but a smaller sound
horizon. This makes the peak separation larger. The Brane model 
has the smallest sound horizon of all, but it
also has the smallest angular diameter distance to the LSS since the
universe is younger. The two roughly cancel each other and give
 roughly the same peak separation as in a $\Lambda$CDM model.
\begin{figure}[h]
\centerline{\hbox{\psfig{file=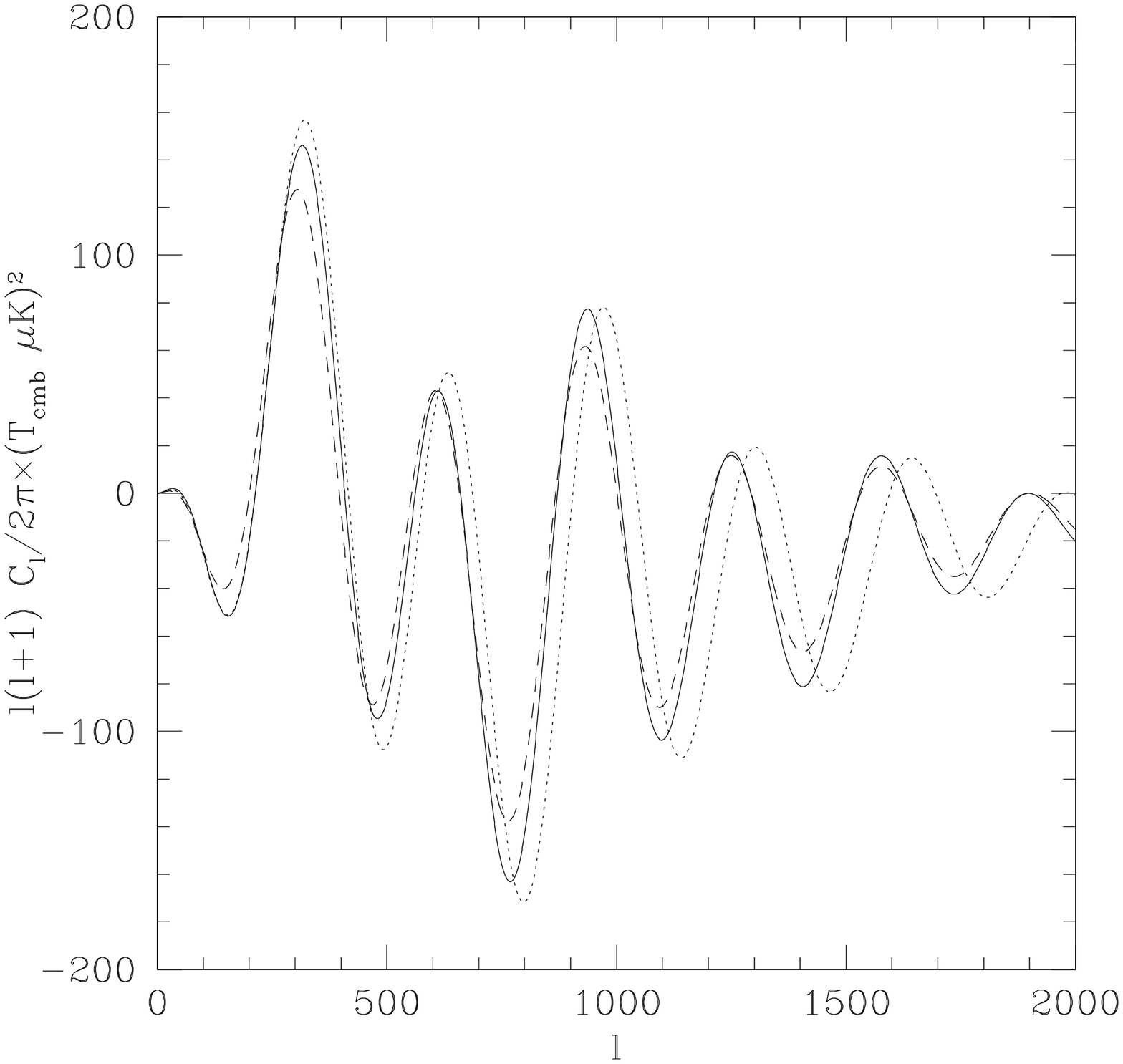,width=3.5in}\hspace{1cm}
\psfig{file=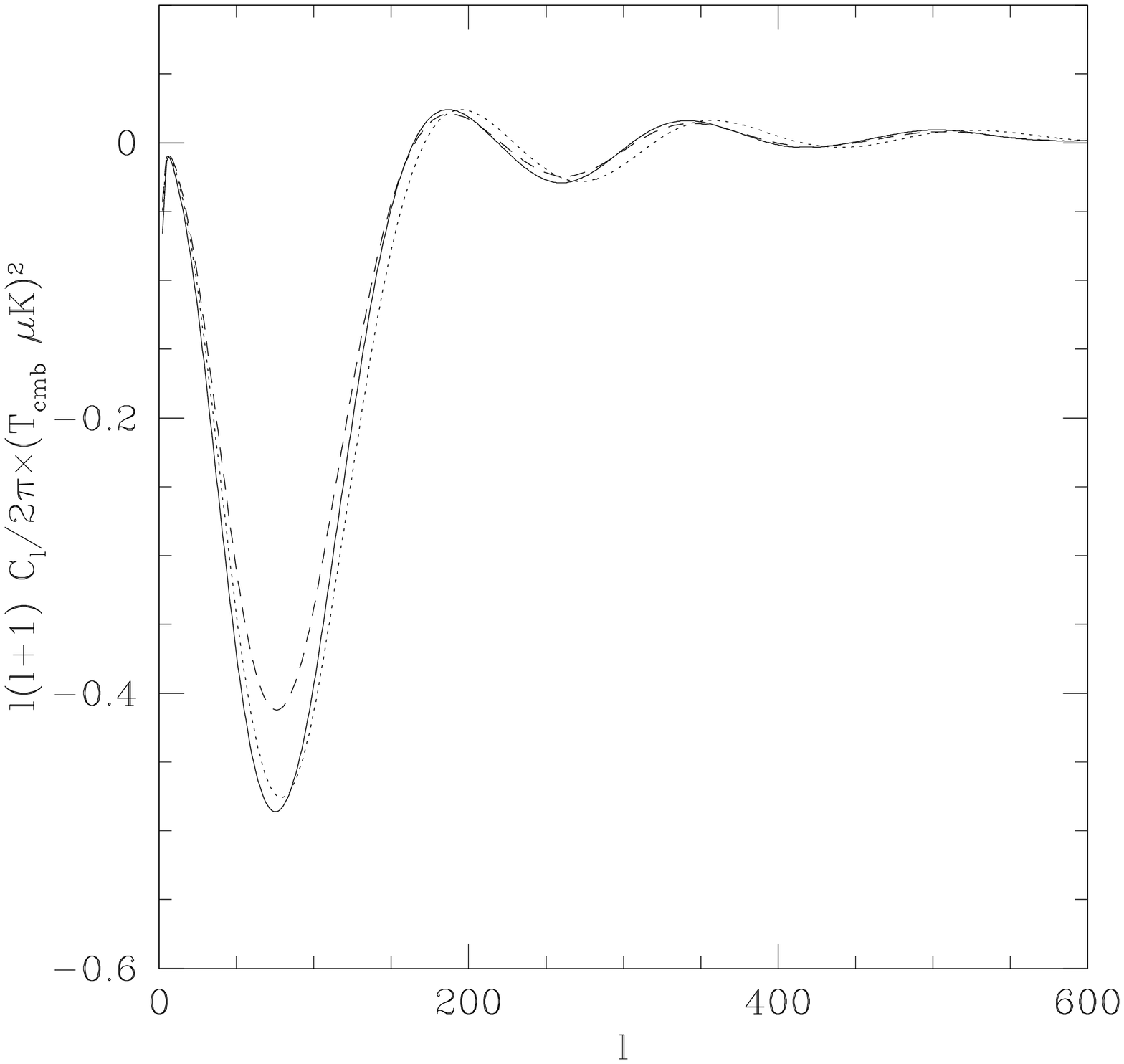,width=3.5in}}}
\caption{Cross correlation of temperature and polarization. On the left we have scalar
cross correlation for the same  models as Fig. \ref{twopotentials}. On the right
we have tensor cross correlation for the same models as Fig.\protect\ref{tensor} In both cases
we see a strong suppression of the correlation amplitude for the Brane model due to the ISW effect.}
\label{cross}
\end{figure}

Then we have the position of the first acoustic peak. The AS model has a first peak
roughly at the same scale as the $\Lambda$ model. The Brane model however has the first peak
shifted to larger scales. This is a combination of two effects. On is the much smaller 
angular diameter distance and the other is the very strong ISW.

\subsection{Tensor modes}
As mentioned previously, the scalar field has no shear and therefore
the only effect it has on the tensor modes is through the background
evolution. The tensor anisotropies are shown in Fig.\ref{tensor}. 

The amplitude of the tensor  modes is in general much less
than the scalar amplitude and the only important contribution comes 
at large scales. Even though they don't affect the temperature 
anisotropy much they are still important as they create B-type polarization  
which might be observed.

The main effect is the strong ISW of the Brane model which suppresses
the anisotropies at very large scales through the mechanism discussed
in the earlier subsection. In the case of the AS model, the ISW is not
as strong, so the model still retains anisotropies of roughly the same 
amplitude as a $\Lambda$ model. Finally we have the shifting of the oscillation
pattern due to the different sound horizon and angular diameter distance just
like the scalar case. 

\subsection{Polarization}
Polarization depends only on the physics at the LSS, unlike the
temperature anisotropy which also depends on the evolution since the
LSS. It is therefore a direct probe of the physics at the LSS. For a review
of polarization see~\cite{HW}
	
 E-type polarization for scalar modes is shown in Fig. \ref{polar}. 
The first feature we observe is a difference in the
position of the peaks. This is due to the same effects as in the scalar 
$C_l$ case, i.e. the sound horizon and angular diameter distance.

The heights of the peaks are affected by the driving effect on the
local quadrupole at LSS. Therefore in the left panel of  Fig. \ref{polar} we see again as expected that
the  AS model has the largest amplitude followed by the $\Lambda$ model. The Brane model
should have had the largest amplitude but we have again the indirect effect of the ISW on
the normalization which suppresses the whole thing.
 This happens again in this case because we normalized to COBE- 
the mechanism of suppression 
being the same as the one described in the scalar $C_l$ case.

B-type polarization is shown on the right of Fig.\ref{polar}. To get a sizable B-type polarization
we used slightly different cosmological parameters(shown in the caption) from the ones used so far, basically changing the
spectral indices and having a non-zero tensor to scalar quadrupole amplitude.  
The difference in the AS and $\Lambda$ models is just in the frequency of oscillation. For the Brane
model we do have however the indirect effect of a strong ISW through normalization of the
temperature anisotropies  amplitude(same thing as in the E-polarization).
 We should note however that lensing can also produce B-type polarization
but we don't discuss it here since in the case of these models quintessence can
 be treated like a cosmological constant as far as lensing is concerned.

Finally we have the polarization-temperature cross correlation shown in
Fig.\ref{cross}. The importance of the cross correlation has been discussed in~\cite{CCT} for scalar
modes and ~\cite{CCT2} for tensor modes.
Again we see the different oscillation frequencies due to
the different sound horizons and angular diameter distances just like the temperature
anisotropies case, for both the scalar and tensor modes.
 In the case of the Brane model the amplitude of the correlation is smaller
simply because in this model the temperature anisotropies (after normalization)
 are smaller compare to the other models. Have we kept the same initial power spectrum for
all models the Brane model would have had the highest correlation. 
In the scalar case, on scales between $l>200$ and $l<1000$ we have the driving 
effect again giving more correlation to the quintessence
models. Moreover for $l<600$ we also have the ISW boosting the correlation for quintessence.
At very small scales the correlation amplitude tends to be the same in
all models. 

\section{The Magnitude-redshift relation}

Before concluding we comment briefly on the magnitude-redshift
relation for the brane model.  The magnitude-redshift relation $m(z)$
for a standard candle of fixed fiducial absolute magnitude is an
essential tool for pinning down the nature of the cosmic
acceleration.  Different quintessence models will produce different
functions $m(z)$ which can be discriminated using data from, for example,
type Ia supernovae (see for example \cite{WellerAlbrecht00}).  When
considering possible future probes of $m(z)$ it is essential to
subject the space of possible models to the constraint that the models
give a realistic account of cosmic structure.  In particular, one can
expect the structure-suppressing nature of quintessence to result in
substantial constraints on the significance of quintessence at $z$
greater than around unity  since structure needs to have a chance 
to form\cite{Turner00}. It is therefor intriguing that that the Brane
model (which gives a good fit to cosmic structure data) has some pretty 
interesting features in $m(z)$.  

\begin{figure}[h]
\centerline{\psfig{file=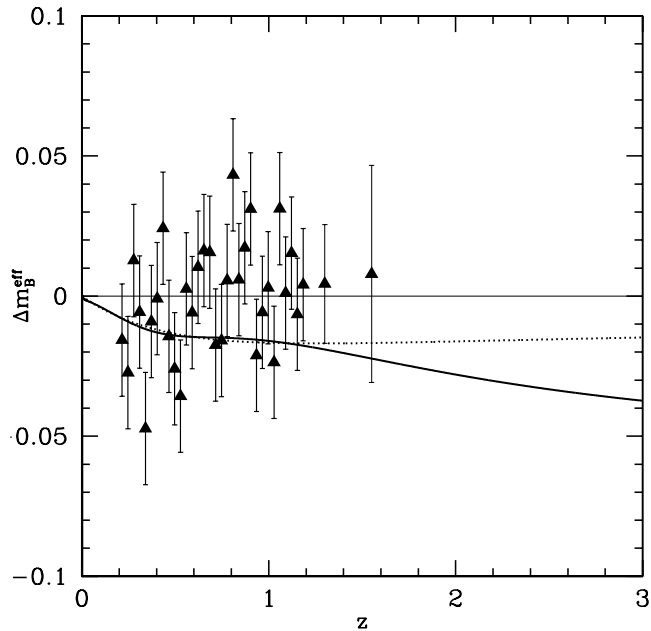,width=3.5in}}
\caption{$\Delta m(z)$ for the Brane model (heavy curve) and a $w=
{\rm const.} = -0.9559 $ model (dashed curve), along with simulated SNAP data.  The
curves are essentially identical up to $z \approx 1.3$ at which point
they diverge. If a standard candle could be found that is effective at
high redshifts it could play a significant role in discriminating
among these types of models (We thank J. Weller for producing this figure).}
\label{mz}
\end{figure}

Figure \ref{mz} shows $\Delta m(z)$ for the
Brane model (identical to the model shown in figure \ref{growth}), a model for
which the dark energy has an equation of state $p=w\rho$ with constant 
w=-0.9559, and simulated binned data representing 1900 supernova
events as might be provided by the proposes SNAP satellite\cite{snap}.  The
$m(z)$ curve for a pure cosmological constant model (which was used to
produce the simulated data) is subtracted from $m(z)$ for the other
models to get $\Delta m(z)$. Except for the choice of theoretical models shown,
Fig. \ref{mz} is identical to  Fig. 2 in \cite{WellerAlbrecht00},
where details of its construction can be found. 

The $\Delta m(z)$ curves for the two dark energy models are
essentially identical up to $z \approx 1.3$ at which point 
they diverge. As discussed in \cite{WellerAlbrecht00,WellerAlbrecht01} there is no
question that a SNAP-type data set will have a high impact on our
ability to discriminate among models of the dark energy. Our work on
the Brane model illustrates how even with a high quality SNAP-class data set
in hand, there could be exciting opportunities to {\em further}
discriminate among realistic models if a standard candle could be
found that is effective out to higher $z$.

\section{Conclusions}

\begin{figure}[h]
\centerline{\hbox{\psfig{file=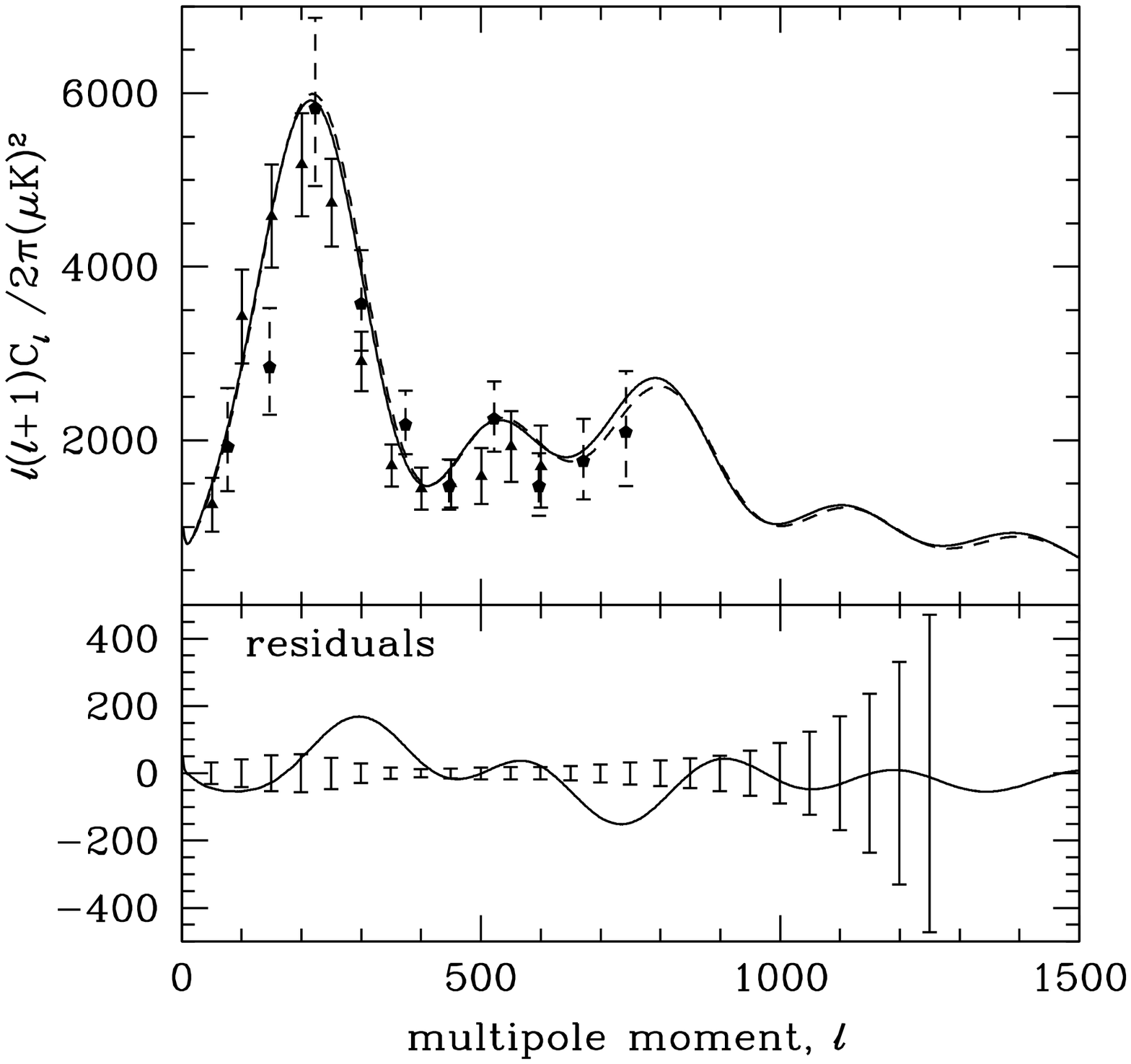,width=3.5in}\hspace{1cm}
\psfig{file=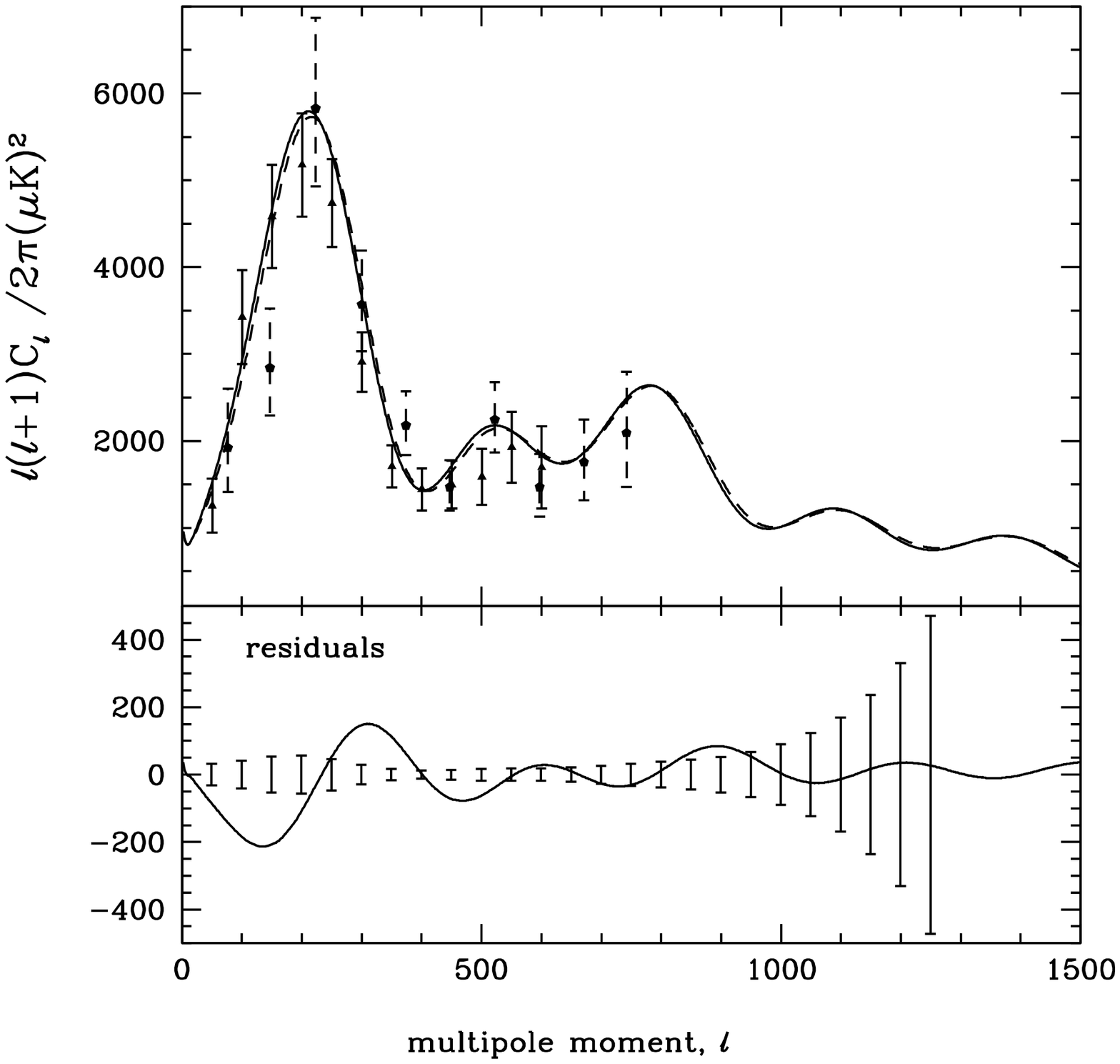,width=3.5in}}}
\caption{A possible future comparison of the AS and Brane models with a $\Lambda$ model using MAP data.
On the left we have the AS model of the right of Fig.\ref{growth} compared with a $\Lambda$ model with $h=0.75$, 
$\Omega_\Lambda = 0.7005$, $\Omega_b = 0.053$, $\Omega_c = 0.2465$ and $\tau=0.1$. On the right
we have the Brane model of the right of Fig.\ref{growth}  compared with a $\Lambda$ model with 
$h=0.75$, $\Omega_\Lambda=0.66327$, $\Omega_b=0.053$, $\Omega_c=0.28373$ and $\tau_{\text{opt}}=0.1$.
 The two $\Lambda$ models were chosen to match the first peak and at the same time the rest of 
the curve as close as possible. On the bottom of each figure we show the residuals of the
 two corresponding models with MAP error-bars (We thank L.Knox for providing the MAP error-bars).}
\label{map}
\end{figure}

We have presented a detailed investigation of a class of quintessence
models motivated by our earlier work\cite{AS}.  These models employ a
particular mechanism (the exponential potential with a prefactor)
which allows realistic models to be produced 
with all potential parameters $O(1)$ in Planck units. The potentials
have a reasonable chance of developing strong theoretical foundations
in brane theory or other theories with extra dimensions.  The work
presented here has taught us that these models also have an
interesting and potentially observable impact on the formation of
cosmic structure. The same mechanism that makes these models
attractive from the point of view of fundamental physics causes the
quintessence to play a much more significant role throughout the
history of the Universe.  This feature leads to interesting effects on the
microwave background anisotropies, the matter power spectrum and the
magnitude redshift relation that result in potentially observable
differences from the predictions of other dark energy models.  We have
extensively examined the physical causes and the nature of 
these effects. 

To illustrate the possibilities, Fig. \ref{map} shows
the full CMB Temperature anisotropy power spectrum for the Brane (right) and
AS (left) models, containing all the effects discussed in this paper.
Each plot also shows the $\Lambda$ model which best mimics the
corresponding quintessence model.  The lower panel of each plot shows
that despite the closeness of the two curves, they are potentially
differentiable by the MAP data (and of course we will eventually have
much more than just the MAP data set).

One issue which still needs to be addressed is the
degree to which the signals we have discovered 
can realistically be differentiated from all possible mimicking
behavior due to the dependence of the predictions on a range of
cosmological parameters (the mimicking $\Lambda$ models in
Fig. \ref{map} were produced ``by hand'', by making a thorough but not 
completely exhaustive exploration of all possible parameters).  This
paper lays the groundwork for such a  project.  The full impact of our
results will not be know until this this issue is addressed in a more
systematic way.  

Still, it is quite interesting that our quintessence models give
predictions that fit all 
existing constraints, and which leave a potentially unique set of
signals that could be observed with future experiments.

\section{Acknowledgments:}
We thank A. Lewin, B. Gold, M. Turner, Y.-S. Song, L.Wang and J. Weller for helpful
conversations. A.A. thanks the Aspen Center for Physics 
and C.S. thanks International Center for Theoretical Physics
where some of this work was completed. This work was supported by DOE grant
DE-FG03-91ER40674 and U.C. Davis.  The CMB and  matter power spectra
used in this work were generated by a modified version of
CMBFAST~\cite{CMBFAST}.


\end{document}